\documentclass[journal]{IEEEtran}
\ifCLASSINFOpdf
\else
\fi
\usepackage{amsmath}
\usepackage{amssymb}
\DeclareMathAlphabet{\pazocal}{OMS}{zplm}{m}{n}
\usepackage{bbm}

\usepackage{float}
\usepackage[export]{adjustbox}
\usepackage{graphicx}
\usepackage[colorlinks=true, allcolors=blue]{hyperref}
\usepackage[utf8]{inputenc}
\usepackage[export]{adjustbox}
\usepackage{caption}
\usepackage{subcaption}
\usepackage{multirow}
\usepackage{multicol}
\usepackage{amsmath}
\usepackage{amsfonts}
\usepackage{bm}
\usepackage{bbm}
\usepackage{algorithmic}
\usepackage[ruled,vlined]{algorithm2e}
\usepackage{booktabs}
\hypersetup{pdfstartview={XYZ null null 1.00}}
\begin{document}
%
\title{Variational-Based Nonlinear Bayesian Filtering with \textcolor{black}{Biased} Observations}
%
%
%

\author{Aamir Hussain Chughtai, Arslan Majal, Muhammad Tahir, \IEEEmembership{Senior Member, IEEE}, and Momin Uppal, \IEEEmembership{Senior Member, IEEE}
	\thanks{The authors are with Department of Electrical Engineering, Lahore University of Management Sciences, DHA Lahore Cantt., 54792, Lahore Pakistan. { (email: aamir.chughtai@lums.edu.pk; arslan.majal@lums.edu.pk; tahir@lums.edu.pk; momin.uppal@lums.edu.pk)}}%
}
%
%

\markboth{Journal of \LaTeX\ Class Files,~Vol.~xx, No.~x, xx~20xx}%
{Shell \MakeLowercase{\textit{et al.}}: Bare Demo of IEEEtran.cls for IEEE Journals}
%



\maketitle

\begin{abstract}
State estimation of dynamical systems is crucial for providing new decision-making and system automation information in different applications. However, the assumptions on the standard computational models for sensor measurements can be violated in practice due to different types of data abnormalities \textcolor{black}{such as outliers and biases}. In this work, we focus on the occurrence of \textcolor{black}{measurement biases} and propose a robust filter for their detection and mitigation during state estimation of nonlinear dynamical systems. We model the presence of \textcolor{black}{bias in each dimension} within the generative structure of the state-space models. Subsequently, employing the theory of Variational Bayes and general Gaussian filtering, we devise a recursive filter which we call the Bias Detecting and Mitigating (BDM) filter. As the error detection mechanism is embedded within the filter structure its dependence on any external detector is obviated. Simulations verify the performance gains of the proposed BDM filter compared to similar Kalman filtering-based approaches in terms of robustness to temporary and persistent bias presence.
\end{abstract}


%
\IEEEpeerreviewmaketitle

\section{Introduction}
State estimation of a dynamical system  plays an indispensable role in the correct functionality of a wide variety of  applications such as energy management systems \cite{9088966}, smart grids \cite{8546788}, robotics \cite{barfoot2017state}, and intelligent transportation systems \cite{7775094}. For linear state-space models (SSMs) with additive White Gaussian noise (AWGN), Kalman Filter (KF) is the optimal state estimator in terms of the minimum mean square error (MSE) criterion \cite{kalman1960new}. For nonlinear SSMs, the filtering becomes more challenging owing to the underlying functional nonlinearities. For such systems, several filtering approaches have been presented in the literature including the Extended Kalman Filter (EKF) \cite{grewal2014kalman}, the Unscented Kalman Filter (UKF)\cite{julier1997new}, Particle Filters (PFs) \cite{gordon1993novel} etc. For a survey of nonlinear state estimators, the reader is referred to \cite{norgaard2000new}.

The performances of standard filtering approaches rely on the assumption that the statistics of \textit{nominal} noise entering the system are perfectly known \textit{apriori}. However, the assumption can easily be violated in practice when the measurements are disturbed by noise not described by the known statistics. \textcolor{black}{In addition to the nominal noise, the observations can further be corrupted by other abnormalities commonly referred to as \textit{gross errors} which include outliers and biases \cite{narasimhan1999data}. Generally modeled with zero-mean statistics, outliers are sudden sporadic surges in the measurements. Their occurrence can be attributed to factors like sensor degradation, communication failures, environmental influences, etc. \cite{9050910}. On the other hand, biases manifest in the data with non-zero mean noise statistics \cite{9050891}. Multiple factors in various applications lead to the appearance of biased observations. Examples include miscalibrations of sensors, other configuration aberrations like errors in sensor location or alignment, clock errors, or malfunctioning \cite{belfadel2019single}. In this work, we keep our focus only on the appearance of \textcolor{black}{measurement biases} and how these can be compensated within the filtering framework.}

Since the magnitude of biases in the data, the instances of their occurrence, and the particular measurement dimensions which get affected are unknown and only partial statistics describing such corruptions are available, the problem of filtering during their presence is nontrivial. The challenge is further compounded by the functional non-linearities appearing in the SSMs. Given the significance of dealing with \textcolor{black}{biases} in the data during filtering, the topic has historically garnered the attention of various investigators. The approaches for devising bias-robust filters differ in the way such anomalies are neutralized in the filtering process.

Conventionally, biases are catered by assuming that the affected measurement dimensions are known in advance. Moreover, the bias evolution models are assumed to be simplistic or completely ignored during filtering. A straightforward approach is to jointly consider the state vector and bias vector for inference supposing the biases are described in a simple Markovian manner. With computational limitations at that time, earlier works attempted to reduce the processing overhead for such formulations \cite{1099223}. In a similar vein, Schmidt aimed to simplify the joint state and bias estimation in the SKF formulation resulting in the celebrated Schmidt Kalman Filter (SKF) \cite{schmidt1966application}. Interestingly, the bias is not estimated at each time step and only its correlations with the state are updated instead, making the SKF \textit{suboptimal} even if the bias transition can perfectly be modeled \cite{grewal2014kalman}. Ideas similar to the SKF have also been proposed to cater for biased measurements, in terms of exploiting partial information e.g. positivity of biases \cite{filtermobile}. These kinds of \textit{suboptimal} approaches are more useful in scenarios where it is safe to ignore the information regarding the evolution of \textcolor{black}{measurement biases}. With the advances in available processing power, the joint state and parameter estimation approach, in the KF framework, remains the standard go-to approach for catering \textcolor{black}{biased observations}\cite{zhang2021extended}.

Conventional methods are more relevant when the bias manifests in the observations consistently throughout the entire duration of system operation. However, these methods inherently assume prior knowledge regarding the affected measurement dimensions. For example, the authors do not consider bias estimation in the angle of arrival (AOA) measurements for filtering \cite{1326705}. Such information may be available beforehand for some applications. However, it must  generally be obtained from some \textit{detection} mechanism for the algorithms to work properly especially if the compromised dimensions vary over time. 

As a result, more sophisticated schemes have been proposed by integrating the detection process with the filtering framework. To this end, following two possibilities exist: 1) use some external/separate detectors, 2) incorporate the detection/compensating process within a unified filtering framework. Both of these approaches have their merits and drawbacks. External detectors are particularly advantageous in terms of their off-the-shelf accessibility to several options. However, the performance of robust filtering is highly dependent on the functionality of these detectors and the way they integrate. On the other hand, internal detection methodologies are harder to design but obviate dependence on external algorithms.      

In the literature, several kinds of external bias detectors sometimes called fault detection and identification algorithms (FDI), have been reported for several applications like tracking using UWB, GPS, and UMTS measurements. For example, in \cite{886790}, biased measurements are identified simply by comparison of the standard deviation of range measurements with a detection threshold. A similar approach is to use normalized residuals to detect the presence of \textcolor{black}{any bias}\cite{hu2020robust}. The method proposed in \cite{562692} uses a historical record of sequential observations and performs a hypothesis test for detection. For bias detection, the use of classical statistical hypothesis tests like likelihood ratio test (LRT) and other probability ratio tests, has also been documented \cite{686556,7080484,6549130, hu2020robust}. In addition, other methods resort to deep learning for the determination of affected measurements \cite{9108193}. There are several bias compensating filtering methods that rely on these kinds of separate detectors \cite{8805384,filtermobile,1583910}.

The other approach based on inherent bias detection and compensation for filtering is more challenging due to two underlying reasons. First, modeling bias inside the SSM, in a Markovian fashion, is tricky. As noted in \cite{jourdan2005monte} the bias transition cannot be simply modeled as a Gaussian centered at the current bias value. In \cite{jourdan2005monte}, the authors model the bias stochastically by assuming it remains clamped to the previous value with certain predefined probability and jumps with the remaining probability within a set range represented as a uniform distribution. In \cite{gonzalez2009mobile}, a similar model for describing bias is used. The use of random bias model has been proposed in \cite{chughtai2020robust} where the bias is represented with Gaussian density and subsequently the mean of the distribution is described in a Markovian manner. The use of Bernoulli random vectors is proposed to switch between multiple models catering for the presence and absence of \textcolor{black}{biases}. These kinds of models describe \textcolor{black}{measurement biases} effectively but make the use of KFs variants difficult since the underlying distributions for modeling do not remain Gaussian. Furthermore, the sole use of Gaussian approximations for inference no longer remains suitable. Therefore, the authors in \cite{jourdan2005monte,gonzalez2009mobile,chughtai2020robust}, opt for the powerful PFs for the inference that can effectively handle arbitrary probability densities. The use of a large number of particles ranging from 15000 to 20000 is reported in these works. Therefore, the computational overhead of PF remains a concern and can be prohibitive in different applications.   

Literature survey indicates the need for computationally efficient methods, with inherent detection mechanisms, to deal with \textcolor{black}{measurement biases}. Though PFs, with reduced processing burden, can be devised and several researchers are concerned with this direction of investigation \cite{1036890,5443441}, an alternate is to use the Kalman filtering results for devising tractable robust filters. Recently, the use of Variational Bayes methods has gained traction in this regard and we focus on this class of approach. Though several robust methods belonging to this class exist that consider \textcolor{black}{outliers in data} \cite{8398426,6349794,8869835,chughtai2021outlier}, however, generally there is a shortage of such methods that deal with \textcolor{black}{biased measurements}. Recent attempts in this direction include the work in \cite{9050891} where the authors consider the Student's-t-inverse-Wishart distribution to handle time-varying bias. However, only linear systems are considered in the derivation.     

Given the backdrop, we present a novel robust filtering method to deal with \textcolor{black}{biases} in the measurements for nonlinear systems. {The main contributions of this work are as follows.}
\begin{itemize}
	\item  {We present the bias detecting and mitigating (BDM) filter to deal with \textcolor{black}{measurements bias} using the Gaussian filtering framework. We resort to Variational inference for designing the filter with an internal bias detection mechanism as opposed to schemes requiring external detectors.} 
	\item  {For a given model we evaluate the Posterior Cramer-Rao Bound (PCRB) to determine the theoretical benchmark for the error performance.}
	\item  {For two different scenarios i.e. persistent and temporary bias presence we evaluate the performance of the BDM filter as compared to different estimators in the literature indicating the advantages of the proposed method.}
\end{itemize}

The way we have organized this article is as follows. Section \ref{modelling_sec} describes our modeling choice for incorporation of \textcolor{black}{biases} inside the SSM. In Section \ref{inference}, the derivation of the filter is provided. Subsequently, the performance evaluation results have been discussed in Section \ref{simulation_sec}. Lastly, conclusive comments are given in Section \ref{Conc}.
\textcolor{black}{\subsection*{Related Works}}
\textcolor{black}{Robust filtering theory has been extensively investigated in the literature. Researchers have been interested in various dimensions of the problem including the design and evaluation of filters robust to different effects like anomalies (outliers and biases) in the process and measurement models, missing observations, system modeling errors, adversarial attacks on sensors, etc. Several robust filters have been proposed and evaluated for different applications resorting to techniques from diverse areas. In this subsection, we briefly highlight some other robust filtering methods.} 

\textcolor{black}{The classical Wiener filter \cite{wiener1949extrapolation} has been successfully extended to its robust counterparts in the literature. The basic idea is to consider the least favorable power spectral densities (PSDs) for any specific uncertainty model assumed for the signal of interest and the noise. Subsequently, the optimal (using min-max MSE criterion) solution is sought to devise such filters \cite{1056875}.} 

\textcolor{black}{Similarly, the KF has numerous robust extensions apart from the ones discussed in the main introduction section. Some of its basic outlier-robust derivatives include $3\sigma$-rejection and score function type KFs \cite{zoubir2018robust}. These formulations use score functions applied to the residuals of the observed and predicted measurements to minimize the effect of outliers. Another robust KF derivative is the approximate conditional mean (ACM) filter which is based on the approximation of the conditional observation density prior to updating \cite{1100882}. Some other variations rely on a bank of KFs to gain robustness \cite{schick1994robust}. The cause of robust filtering has also been well served by the theory of M-estimation which has helped develop many of these methods \cite{zoubir2018robust}. In addition, regression-based KFs have also been proposed for achieving robustness \cite{5371933}. Similarly, guaranteed cost-based methods \cite{317138}, Krein space methods \cite{lee2004robust} and linearly constrained KFs (LCKFs) have also been proposed in this regard \cite{9638328}. Ambiguity sets for catering model distributional uncertainties have also been employed to this end \cite{shafieezadeh2018wasserstein}. Besides different information theoretic criteria have been used to devise robust KFs \cite{chen2017maximum}. KFs extensions to deal with system parametric uncertainties are also well-documented \cite{lewis2017optimal}.}

\textcolor{black}{Other types of robust filtering approaches have also been reported in the literature. $\text{H}_{\infty}$ is a popular approach that aims to minimize the worst-case estimation error by formulating a min-max problem using a smartly chosen objective function \cite{simon2006optimal}. Similarly, mixed Kalman/$\text{H}_{\infty}$ approaches have also been proposed leveraging the merits of both the methods \cite{7555348}. The use of finite impulse response (FIR) filters can also be found in this regard \cite{5428832,8355704,8744320}. Similarly, robust recursive estimators for SSMs resorting to sensitivity penalization-based methods have been devised \cite{zhou2010sensitivity}. The use of nonparametric techniques can also be found in the robust filtering literature as well \cite{zoubir2018robust}.} \\


\section{\textcolor{black}{Bias} Modeling}\label{modelling_sec}
\textcolor{black}{As the standard SSM does not consider the possibility of \textcolor{black}{measurement biases} in its \textit{generative} structure \cite{6266757} it needs to be modified}. At the same time, the model should remain amenable for VB inference. To this end, we choose the inference model from our previous work \cite{chughtai2020robust}, with a few modifications. For a discrete time SSM, the process and measurement equations are given as follows 
\begin{align}
	\mathbf{x}_k= & \mathbf{f}(\mathbf{x}_{k\text{-}1})+\mathbf{q}_{k\text{-}1}\label{eqn1}\\
	\mathbf{y}_k =& \mathbf{h}(\mathbf{x}_{k})+\mathbf{r}_k + \boldsymbol{{\mathcal{I}}}_{k}{\mathbf{\Theta}}_k \label{eqn2}
\end{align}
where $k$ denotes the time-index,  $\mathbf{x}_k\in\mathbb{R}^n$ and $\mathbf{y}_k\in\mathbb{R}^m$ are the state and measurement vectors respectively, $\mathbf{q}_{k\text{-}1}\in\mathbb{R}^n$ and $\mathbf{r}_k\in\mathbb{R}^m$ are white process and  measurement noise vectors, $\mathbf{f}(.)$ and $\mathbf{h}(.)$ represent nonlinear process and measurement dynamics respectively, $\mathbf{\Theta}_k\in\mathbb{R}^m$ models the effect of biases in the measurements and $\boldsymbol{\mathcal{I}}_{k} \in \mathbb{R}^{m\times m}$ is a diagonal matrix with Bernoulli elements ${\mathcal{I}}^i_{k}$ used to indicate the occurrence of bias in different dimensions. We assume the following noise distributions: $\mathbf{q}_{k\text{-}1} \sim \mathcal{N}(\mathbf{0},\mathbf{Q}_{k\text{-}1})$ and $\mathbf{r}_k  \sim \mathcal{N}(\mathbf{0}, \mathbf{R}_k)$. We assume that measurements are obtained from independent sensors making $\mathbf{R}_k$ diagonal. For inferential tractability, the model as originally reported is simplified by ignoring the added randomness in the bias magnitude in \eqref{eqn2}. The bias evolution is expressed as follows where the modeling rationale remains the same as originally reported in \cite{chughtai2020robust}.
\begin{equation}
		{\mathbf{\Theta}}_k = (\mathbf{I}-\boldsymbol{{\mathcal{I}}}_{k\text{-}1}){\widetilde{\mathbf{\Theta}}}_{k} +\boldsymbol{{\mathcal{I}}}_{k\text{-}1}({\mathbf{\Theta}}_{k\text{-}1} +  {\Delta_k})\label{eqn3}
\end{equation}
 In \eqref{eqn3}, each entry of ${\Delta_k}$ allows for any drifts/changes in the bias value over time, in the corresponding dimension, given bias was present at the previous time step. On the contrary, if no bias occurred in any given dimension, at the preceding instant, it can possibly occur with a very large variance ${\sigma^2_{\widetilde{{\Theta}}}}$ (assuming an uninformative prior) described by the respective entries of a zero mean random vector ${\widetilde{\mathbf{\Theta}}}_{k}$. The distributions of ${\Delta_k}$ and ${\widetilde{\mathbf{\Theta}}}_{k}$ are supposed to be white and normally distributed given as
\begin{align}
	{\Delta_k} &\sim \mathcal{N}(\mathbf{0},\breve{\mathbf{\Sigma}}_{k}) \text{ with } \breve{\mathbf{\Sigma}}_{k}= \mathrm{diag}\left(\sigma^2_{\vartriangle{1}}, \cdots, \sigma^2_{\vartriangle{m}} \right) \label{PPFeqn30}&\\
	{\widetilde{\mathbf{\Theta}}}_{k} &\sim \mathcal{N}(\mathbf{0},\widetilde{\mathbf{\Sigma}}_{k}) \text{ with } \widetilde{\mathbf{\Sigma}}_{k} = \mathrm{diag}\left({\sigma^2_{\widetilde{{\Theta}}}}, \cdots, {\sigma^2_{\widetilde{{\Theta}}}} \right) \label{PPFeqn31}
\end{align}
Note that for tractability, we have modified ${\widetilde{\mathbf{\Theta}}}_{k}$ to be normally distributed instead of obeying a uniform distribution. However, assuming a very large variance does not make a practical difference. Also note that for simplicity, we do not take any transition model for $\boldsymbol{\mathcal{I}}_{k}$ and assume its elements occur independently at each instance. Remaining modeling assumptions are kept the same as originally reported. 
\section{Recursive Bayesian Inference}\label{inference}
Considering the inference model in \eqref{eqn1}-\eqref{eqn3}, the Bayes rule can be employed recursively to express the joint posterior distribution of $\mathbf{x}_k$, $\bm{\mathcal{I}}_k$ (considering only the random entries ${\mathcal{I}}^i_k$) and ${\mathbf{\Theta}}_k$ conditioned on the set of all the observations $\mathbf{y}_{1:{k}}$ analytically as \begin{equation}
	p(\mathbf{x}_k,\bm{\mathcal{I}}_k,{\mathbf{\Theta}}_k|\mathbf{y}_{1:{k}})=\frac{p(\mathbf{y}_k|\mathbf{x}_{k},\bm{\mathcal{I}}_k,{\mathbf{\Theta}}_k)	p(\mathbf{x}_k,\bm{\mathcal{I}}_k,{\mathbf{\Theta}}_k|\mathbf{y}_{1:{k\text{-}1}})}{p(\mathbf{y}_k|\mathbf{y}_{1:{k\text{-}1}})}
	\label{eqn_vb_1}
\end{equation} 
Theoretically, the joint posterior can be marginalized to obtain the expression for $p(\mathbf{x}_k|\mathbf{y}_{1:{k}})$. With this approach, the exact sequential Bayesian processing becomes computationally infeasible. Therefore, we adopt the VB method \cite{vsmidl2006variational} for inference where the product of VB marginals is conveniently used to approximate the joint posterior as
\begin{equation}
	p(\mathbf{x}_k,\bm{\mathcal{I}}_k,{\mathbf{\Theta}}_k|\mathbf{y}_{1:{k}})\approx q(\mathbf{x}_k)q(\bm{\mathcal{I}}_k)q({\mathbf{\Theta}}_k)
	\label{eqn_vb_2}
\end{equation} 
With an objective to minimize the Kullback-Leibler divergence (KLD) between the marginal product and the true posterior, the VB method leads to the following marginals 
\begin{align}
	q(\mathbf{x}_k)&\propto \exp( \big\langle\mathrm{ln}(p(\mathbf{x}_k,\bm{\mathcal{I}}_k,{\mathbf{\Theta}}_k|\mathbf{y}_{1:{k}})\big\rangle_{ q({{\bm{\mathcal{I}}}_k}) {q(\mathbf{\Theta}}_k)})\label{eqn_vb_3}\\
	q(\bm{\mathcal{I}}_k)&\propto \exp ( \big\langle\mathrm{ln}(p(\mathbf{x}_k,\bm{\mathcal{I}}_k,{\mathbf{\Theta}}_k|\mathbf{y}_{1:{k}})\big\rangle_{{q(\mathbf{x}_k){q(\mathbf{\Theta}}_k)}})\label{eqn_vb_4}\\
	q({\mathbf{\Theta}}_k)&\propto \exp ( \big\langle\mathrm{ln}(p(\mathbf{x}_k,\bm{\mathcal{I}}_k,{\mathbf{\Theta}}_k|\mathbf{y}_{1:{k}})\big\rangle_{q(\mathbf{x}_k)q(\bm{\mathcal{I}}_k)})\label{eqn_vb_5}
\end{align} 
where $ \langle.\rangle_{q(\bm{\psi}_k)}$ denotes the expectation of the argument with respect to a distribution $q(\bm{\psi}_k)$. The VB marginals can be updated iteratively until convergence, using \eqref{eqn_vb_3}-\eqref{eqn_vb_5} in turn. The procedure provides a convenient way to approximate the true marginals of the joint posterior by approximating these as $ p(\mathbf{x}_k|\mathbf{y}_{1:{k}})$$\approx$$ {q^c(\mathbf{x}_k)}$, $	p(\bm{\mathcal{I}}_k|\mathbf{y}_{1:{k}})$$\approx$${q^c(\bm{\mathcal{I}}_k)}$ and $	p({\mathbf{\Theta}}_k|\mathbf{y}_{1:{k}})$$\approx$$ 	{q^c({\mathbf{\Theta}}_k)}$ where ${q^c(.)}$ denotes the VB marginals obtained after convergence.

\subsection{Prediction}
Assuming that at each time step the posterior is approximated with a product of marginals, the predictive density can be \textcolor{black}{approximated} as
\begin{align}
&p(\mathbf{x}_k,\bm{\mathcal{I}}_k,{\mathbf{\Theta}}_k|\mathbf{y}_{1:{k\text{-}1}})\approx p(\bm{\mathcal{I}}_k)p(\mathbf{x}_k|\mathbf{y}_{{1:k\text{-}1}}) p({\mathbf{\Theta}}_k|\mathbf{y}_{1:k\text{-}1}) \label{eqn_vb_8}
\end{align} 
with 
\begin{align}
	p(\mathbf{x}_k|\mathbf{y}_{1:{k\text{-}1}})&= \int p(\mathbf{x}_k|\mathbf{x}_{k\text{-}1}) p(\mathbf{x}_{k\text{-}1}|\mathbf{y}_{1:{k\text{-}1}}) d\mathbf{x}_{k\text{-}1} \label{eqn_vb_9}\\
	p({\mathbf{\Theta}}_k|\mathbf{y}_{1:k\text{-}1}) &\approx \int\int p({\mathbf{\Theta}}_k|\bm{\mathcal{I}}_{k\text{-}1},{\mathbf{\Theta}}_{k\text{-}1}) p(\bm{\mathcal{I}}_{k\text{-}1}|\mathbf{y}_{1:{k\text{-}1}})\nonumber\\ &\hspace{.5cm}p({\mathbf{\Theta}}_{k\text{-}1}|\mathbf{y}_{1:{k\text{-}1}}) d\bm{\mathcal{I}}_{k\text{-}1}d{\mathbf{\Theta}}_{k\text{-}1}\label{eqn_vb_10}
\end{align}
We assume that the occurrence of bias is independent for each dimension and its historical existence. Using $\theta^i_k$ to denote the prior probability of occurrence of bias in the $i$th observation, the distribution of $\bm{\mathcal{I}}_k$ is defined as product of independent Bernoulli distributions of each element
\begin{equation}
	p(\bm{\mathcal{I}}_k)=\prod_{i=1}^{m}p({{\mathcal{I}}}^i_k)=\prod_{i=1}^{m} (1-{\theta^i_k}) \delta({{{\mathcal{I}}}^i_k})+{\theta^i_k}\delta( {{{\mathcal{I}}}^i_k}-1)
	\label{eqn_model_11}
\end{equation}
where $\delta(.)$ denotes the delta function.

To evaluate \eqref{eqn_vb_9}-\eqref{eqn_vb_10}, suppose the following distributions for the posterior marginals at time step $k-1$ 
\begin{align}
	&p(\mathbf{x}_{k\text{-}1}|\mathbf{y}_{1:{k\text{-}1}})\approx {q^c(\mathbf{x}_{k\text{-}1})}\approx \mathcal{N}(\mathbf{x}_{k\text{-}1}|\mathbf{\hat{x}}^+_{k\text{-}1},\mathbf{{P}}^+_{k\text{-}1}) \label{eqn_vb_11}\\
	&p(\bm{\mathcal{I}}_{k\text{-}1}|\mathbf{y}_{1:{k\text{-}1}})\approx {q^c(\bm{\mathcal{I}}_{k\text{-}1})}=\prod_{i=1}^{m}p({{\mathcal{I}}}^i_{k\text{-}1}|\mathbf{y}_{1:{k\text{-}1}})\nonumber \\
	&=\prod_{i=1}^{m} (1-{\Omega^i_{k\text{-}1}}) \delta({{{\mathcal{I}}}^i_{k\text{-}1}})+{\Omega^i_{k\text{-}1}}\delta( {{{\mathcal{I}}}^i_{k\text{-}1}}-1) \label{eqn_vb_12}	\\
	&p({\mathbf{\Theta}}_{k\text{-}1}|\mathbf{y}_{1:{k\text{-}1}})\approx {q^c({\mathbf{\Theta}}_{k\text{-}1})} \approx\mathcal{N}(\mathbf{\Theta}_{k\text{-}1}|\mathbf{\hat{\Theta}}^+_{k\text{-}1},\mathbf{{\Sigma}}^+_{k\text{-}1})\label{eqn_vb_13}
\end{align}
where ${\Omega^i_{k}}$ denotes the posterior probability of bias occurrence in the $i$th dimension. The notation $\mathcal{N}(\mathbf{x}|\mathbf{m},\mathbf{{\Sigma}})$ represents a multivariate normal distribution with mean $\mathbf{m}$ and covariance $\mathbf{{\Sigma}}$, evaluated at $\mathbf{x}$. The verification of the functional forms of the distributions and the expressions of their parameters are provided in the subsequent update step of the Bayesian filter.

Since $\mathbf{f}(.)$ is assumed to \textcolor{black}{be} nonlinear, $p(\mathbf{x}_k|\mathbf{y}_{1:k\text{-}1})$ can be approximated, using general Gaussian filtering results \cite{sarkka2013bayesian}, as $\mathcal{N}(\mathbf{x}_{k}|\mathbf{\hat{x}}^-_{k},\mathbf{{P}}^-_{k})$ with the parameters predicted as follows 
\begin{flalign}
	\mathbf{\hat{x}}^{-}_{k}&=\big\langle \mathbf{f}(\mathbf{x}_{k\text{-}1})\big\rangle_{p(\mathbf{x}_{k\text{-}1}|\mathbf{y}_{1:k\text{-}1})}&\label{eqn_vb_14}\\
	\mathbf{P}^{-}_{k}&=\big\langle(\mathbf{f}(\mathbf{x}_{k\text{-}1})-\mathbf{\hat{x}}^{-}_{k})(\mathbf{f}(\mathbf{x}_{k\text{-}1})-\mathbf{\hat{x}}^{-}_{k})^{\top}\big\rangle_{p(\mathbf{x}_{k\text{-}1}|\mathbf{y}_{1:{k\text{-}1}})}+\mathbf{Q}_{k\text{-}1}\label{eqn_vb_15}
\end{flalign}
The remaining term required to approximate the predictive density recursively in \eqref{eqn_vb_8} is $p({\mathbf{\Theta}}_k|\mathbf{y}_{k\text{-}1})$. Observing \eqref{eqn3} and \eqref{eqn_vb_10}, it is evident that $p({\mathbf{\Theta}}_k|\mathbf{y}_{k\text{-}1})$ is a sum of $2^m$ Gaussian densities scaled by the probabilities of combinations of bias occurrence at previous time instance. Obviously this makes recursive inference intractable, so we propose to approximate this distribution with a single Gaussian density $\mathcal{N}(\mathbf{\Theta}_{k}|\mathbf{\hat{\Theta}}^-_{k},\mathbf{{\Sigma}}^-_{k})$ using moment matching \cite{sarkka2013bayesian}. The parameters of the distribution are updated as
\begin{flalign}
	\mathbf{\hat{\Theta}}^{-}_{k}&=\mathbf{\Omega}_{k\text{-}1}\mathbf{\hat{\Theta}}^{+}_{k\text{-}1}\label{eqn_vb_16}\\
	\mathbf{\Sigma}^{-}_{k}&=(\mathbf{I}-\mathbf{\Omega}_{k\text{-}1}) \widetilde{\mathbf{\Sigma}}_{k}+\mathbf{\Omega}_{k\text{-}1}\breve{\mathbf{\Sigma}}_{k}& \nonumber\\
	&+\mathbf{{\Sigma}}^+_{k\text{-}1}\odot(\mathrm{diag}(\mathbf{\Omega}_{k\text{-}1}){\mathrm{diag}(\mathbf{\Omega}_{k\text{-}1})}^{\top}+\mathbf{\Omega}_{k\text{-}1}(\mathbf{I}-\mathbf{\Omega}_{k\text{-}1}))&\nonumber \\
	&+\mathbf{\Omega}_{k\text{-}1}(\mathbf{I}-\mathbf{\Omega}_{k\text{-}1})(\mathrm{diag}(\mathbf{\hat{\Theta}}^{+}_{k\text{-}1}))^2 &	
	\label{eqn_vb_17}
\end{flalign}
where $\mathbf{\Omega}_{k\text{-}1}$ is a diagonal matrix with entries ${\Omega}^i_{k\text{-}1}$ denoting the posterior probability of bias occurrence at time step $k-1$. The operator $\odot$ is the Hadamard product and $\mathrm{diag}$ is used for vector to diagonal matrix conversion and vice versa. The reader is referred to the Appendix for detailed derivations of \eqref{eqn_vb_16}-\eqref{eqn_vb_17}.
\subsubsection*{Remarks}
We note the following in \eqref{eqn_vb_16}-\eqref{eqn_vb_17}
\begin{itemize}
	\item $\mathbf{\Omega}_{k\text{-}1}$ dictates how the parameters $\mathbf{\hat{\Theta}}^-_{k}$ and $\mathbf{{\Sigma}}^-_{k}$ are predicted.
	\item For the case when $\mathbf{\Omega}_{k\text{-}1}=\mathbf{I}$, i.e. bias is inferred in each dimension at time step $k{-}1$ with probability 1, $	\mathbf{\hat{\Theta}}^{-}_{k}=\mathbf{\hat{\Theta}}^{+}_{k\text{-}1}$ and  $\mathbf{\Sigma}^{-}_{k}=\mathbf{{\Sigma}}^+_{k\text{-}1}+\breve{\mathbf{\Sigma}}_{k}$. In other words, the mean of the bias prediction (for each dimension) is retained and its covariance is predicted as sum of previous covariance and the covariance considered for the amount of drift/change in the bias.
	\item For the case when $\mathbf{\Omega}_{k\text{-}1}=\mathbf{0}$, i.e. no bias is inferred in each dimension at time step $k{-}1$ with probability 1, $	\mathbf{\hat{\Theta}}^{-}_{k}=\mathbf{0}$ and  $\mathbf{\Sigma}^{-}_{k}=\widetilde{\mathbf{\Sigma}}_{k}$. In other words, the mean of the bias prediction (for each dimension) is $\mathbf{0}$ and its covariance is predicted with very large entries.
	\item Similarly, the prediction mechanism can be understood if only some dimensions are inferred to be disturbed with probability 1. The bias in the particular dimensions are predicted with the mean retained and covariance updated as addition of previous covariance plus the covariance allowed for the drift/change.
	\item Lastly, if there is partial confidence on the occurrence of bias in any dimension at $k{-}1$, the predicted Gaussian distribution is shifted to the mean of bias estimate at $k{-}1$ scaled with a factor of ${\Omega}^i_{k\text{-}1}$. In addition, the covariance gets inflated by addition of scaled elements of $\widetilde{\mathbf{\Sigma}}_{k}$ and squared terms of mean at $k{-}1$. In other words, it can be interpreted in a sense that unless there is a very high confidence of occurrence of bias at the previous time instance, the bias would be predicted with a large covariance.       
\end{itemize}

\subsection{Update}
For the update step, we resort to \eqref{eqn_vb_1}-\eqref{eqn_vb_5} and use \eqref{eqn_vb_8} for approximating the predictive density. For detailed derivations, the reader is referred to the Appendix.

Parameters of $q(\mathbf{x}_k)$ are updated iteratively as
\begin{flalign}
	\hat {\mathbf{x}} _{k} ^{+} &= \hat {\mathbf{x}}_{k}^{-} + \mathbf{K}_k(\mathbf{y}_{k} - {\boldsymbol{\Omega}} _{k} \mathbf{\hat{\Theta}}^+_{k} - \bm{\mu}_k) &\label{eqn_vb_18} \\
	\bm{\mu}_k&=\langle \mathbf{h}(\mathbf{x}_{k})  \rangle_{p(\mathbf{x}_k|\mathbf{y}_{k\text{-}1})}& \label{eqn_vb_19}\\
	\mathbf{P}_{k}^{+} &= \mathbf{P}_{k}^{-} - \mathbf{C}_{k}\mathbf{K}_{k}^{\top}& \label{eqn_vb_20}\\
	\mathbf{K}_{k} &= \mathbf{C}_{k}\mathbf{S}_{k}^{-1}& \label{eqn_vb_21}\\
	\mathbf{C}_{k}&=\big \langle(\mathbf{x}_{k} - \hat {\mathbf{x}}_{k}^{-} )(\mathbf{h}(\mathbf{x}_{k}) - \bm{\mu}_k) \rangle_{p(\mathbf{x}_k|\mathbf{y}_{k\text{-}1})}& \label{eqn_vb_22}\\
	\mathbf{S}_{k}&=\big \langle(\mathbf{h}(\mathbf{x}_{k}) - \bm{\mu}_k)(\mathbf{h}(\mathbf{x}_{k}) - \bm{\mu}_k)^{\top}  \rangle_{p(\mathbf{x}_k|\mathbf{y}_{k\text{-}1})} + \mathbf{R}_{k}& \label{eqn_vb_23}
\end{flalign}

Parameters of ${q(\bm{\mathcal{I}}_k)}$ are updated iteratively as
\begin{flalign}
	\Omega_{k}^{i} &= {\Pr({\mathcal{I}}_{k}^{i} = 1)}/({{\Pr}({\mathcal{I}}_{k}^{i} = 1) + \Pr({\mathcal{I}}_{k}^{i} = 0)})&\label{eqn_vb_29}
\end{flalign}
where denoting $k_1$ as the proportionality constant
\begin{flalign}
&\Pr({\mathcal{I}}^{i}_{k} = 0) = k_1 (1 - \theta_{k}^{i}) \exp{\big({-}\frac {1}{2} \big(\frac {(y^{i}_{k} - {\nu}^i_k )^{2}}{R_{k}^{i}} + \bar{h}^2_k\big)\big)}& \label{eqn_vb_24}\\
&\Pr({\mathcal{I}}_{k}^{i} = 1) = k_1 \theta_{k}^{i} \exp{ \big({-}\frac{1}{2} \frac{\bar{h}^2_k + \bar{\Theta}^2_k + ( {\nu}^i_k + \hat{\Theta}_{k}^{+^i} - y_{k}^{i})^{2}}{R_{k}^{i}}\big)}&\label{eqn_vb_25}\\
&\bm{\nu}_k= \langle {\mathbf{h}}(\mathbf{x}_{k})  \rangle_{q(\mathbf{x}_k)}&\label{eqn_vb_26}
\end{flalign}
\begin{flalign}
&\bar{h}^2_k = \langle(h^{i}(\textbf{x}_{k}) - {\nu}^i_k )^{2}\rangle_{q(\mathbf{x}_k)}& \label{eqn_vb_27}\\
&\bar{\Theta}^2_k = \langle({\Theta}_{k}^{i} - {{\hat{\Theta}}_k^{+i}})^2\rangle_{q(\mathbf{\Theta}_k)}&\label{eqn_vb_28}
\end{flalign}

Parameters of $q(\mathbf{\Theta}_k)$ are updated iteratively as
\begin{flalign}
	\hat {\mathbf{\Theta}}_{k}^{*} &= \hat{\mathbf{\Theta}}_{k}^{-} + \bm{\mathcal{K}}_{k}(\mathbf{y}_{k} - ( \bm{\nu}_k + \boldsymbol{\Omega}_{k} \hat{\mathbf{\Theta}}_{k}^{-}))&\label{eqn_vb_30}\\
	\mathbf{{\Sigma}}^*_{k} &= \mathbf{{\Sigma}}^-_{k} - \bm{\mathcal{C}}_{k}\bm{\mathcal{K}}_{k}^{\top}&\label{eqn_vb_31}\\
	\bm{\mathcal{K}}_{k} &= \bm{\mathcal{C}}_{k}\bm{\mathcal{S}}_{k}^{-1}&\label{eqn_vb_32}\\
	\bm{\mathcal{C}}_{k}&=\mathbf{{\Sigma}}^-_{k}\boldsymbol{\Omega}_{k}^{\top}&\label{eqn_vb_33}
	\\
	\bm{\mathcal{S}}_{k} &= \boldsymbol{\Omega}_{k}\mathbf{{\Sigma}}^-_{k}\boldsymbol{\Omega}_{k}^{\top} + \mathbf{R}_{k}&\label{eqn_vb_34}\\
	\hat{\mathbf{\Theta}}_{k}^{+}&=\mathbf{\Sigma}_{k}^{+}{\mathbf{{\Sigma}}^*_{k}}^{-1}\hat{\mathbf{\Theta}}_{k}^{*}&\label{eqn_vb_35}\\
	\mathbf{\Sigma}_{k}^{+} &= \big(\boldsymbol{\Omega}_{k}(\mathbf{I}-\boldsymbol{\Omega}_{k})\mathbf{R}_{k}^{-1} + {\mathbf{\Sigma}_{k}^*}^{-1}\big)^{-1}&\label{eqn_vb_36}
\end{flalign}

\begin{algorithm}[ht!]
	\SetAlgoLined
	Initialize\ $\hat {\mathbf{x}} _{0} ^{+},\mathbf{P}^{+}_0,\hat{\mathbf{\Theta}}_{0}^{+},
	\mathbf{\Sigma}_{0}^{+}
	,\widetilde{\mathbf{\Sigma}}_{k},\breve{\mathbf{\Sigma}}_{k},\mathbf{Q}_k,\mathbf{R}_k$;
	
	\For{$k=1,2...K$}{
		Evaluate $\hat {\mathbf{x}} _{k}^{-},\mathbf{P}^{-}_k$ with \eqref{eqn_vb_14} and \eqref{eqn_vb_15}\;
		Evaluate $\mathbf{\hat{\Theta}}^{-}_{k},\mathbf{\Sigma}^{-}_{k}$ with \eqref{eqn_vb_16} and \eqref{eqn_vb_17}\;
		Initialize $\theta^i_k$, $\Omega^{i(0)}_k$,$\hat {\mathbf{\Theta}} _{k} ^{+(0)}$ the convergence threshold $\tau$, \textcolor{black}{$\gamma$}$=\tau+1,\text{the iteration index} \ l=1$\;
		Evaluate $\hat {\mathbf{x}} _{k} ^{+(0)}$ and  ${\mathbf{P}^{+ (0)}_k}$ with \eqref{eqn_vb_18}-\eqref{eqn_vb_23}\;
		\While{\textcolor{black}{$\gamma$}$>\tau$}{
			Update $\Omega^{i(l)}_k\ \forall\ i$ with \eqref{eqn_vb_29}-\eqref{eqn_vb_28}\;
			Update $\hat {\mathbf{\Theta}} _{k} ^{+(l)}$ and  ${\mathbf{\Sigma}^{+ (l)}_k}$ with \eqref{eqn_vb_30}-\eqref{eqn_vb_36}\;
			Update $\hat {\mathbf{x}} _{k} ^{+(l)}$ and  ${\mathbf{P}^{+ (l)}_k}$ with \eqref{eqn_vb_18}-\eqref{eqn_vb_23}\;
			Evaluate \textcolor{black}{$\gamma$}$={\|\hat {\mathbf{x}} _{k} ^{+(l)}-{\mathbf{x}} _{k} ^{+(l-1)} \|}/{\|{\mathbf{x}} _{k} ^{+(l-1)}\|}$\;
			$l=l+1$\;
		}
		$\hat {\mathbf{x}} _{k} ^{+}=\hat {\mathbf{x}} _{k} ^{+(l-1)}$ and ${\mathbf{P}^{+}_k}={\mathbf{P}^{+(l-1)}_k}$\;
		$\hat{\mathbf{\Theta}}_{k}^{+}=\hat{\mathbf{\Theta}}_{k}^{+(l-1)}$and $\mathbf{\Sigma}_{k}^{+}=\mathbf{\Sigma}_{k}^{+(l-1)}$\;
	}
	\caption{The proposed BDM filter}
	\label{Algo1}
\end{algorithm} 
\subsection{BDM Filter}
Using the proposed approximations, in the prediction and update steps, we have devised a recursive filter referred as the BDM filter. \textcolor{black}{Unless real-world experiments reveal any prior information regarding the occurrence of bias in each dimension, we propose using an uninformative prior for ${\mathcal{I}}^i_{k}\ \forall\ i$ which is commonly adopted for such cases. The Bayes-Laplace and the maximum entropy methods for obtaining uninformative prior for a parameter with finite values lead to the uniform prior distribution \cite{martz199414,turkman2019computational}. We adopt this choice of prior for our case i.e. $\theta^i_k=0.5\ \forall\ i$ which has been advocated in the literature for designing robust filters \cite{8869835,chughtai2021outlier}.} Algorithm \ref{Algo1} outlines the devised BDM filter.



\section{Numerical Experiments}\label{simulation_sec}
To evaluate the comparative performance of the devised algorithm, numerical experiments have been conducted on an Apple MacBook Air with a 3.2 GHz M1 Processor and 8 GBs of unified RAM using Matlab R2021a.

For comparative fairness, we consider methods based on the Unscented Kalman Filter (UKF) as their basic algorithmic workhorse. The following filters have been taken into account for comparisons: the standard UKF, the selective observations rejecting UKF (SOR-UKF) \cite{chughtai2021outlier}, the Unscented Schmidt Kalman Filter (USKF)\cite{stauch2015unscented} and the constrained Unscented Kalman Filter (CUKF). The CUKF is devised by modifying the CSRUKF \cite{filtermobile} by using the standard UKF instead of SRUKF as its core algorithm. 

In terms of handling data corruption, the standard UKF has no bias compensation means in its construction. By contrast, the SOR-UKF is an outlier-robust filter, with inherent data anomaly detection mechanism, which discards the observations found to be corrupted. Lastly, the USKF and CUKF both compensate for the bias partially and require an external bias detection mechanism. Bias mitigating filters with inherent detection mechanism are generally scant in the literature and most of these are based on the PFs. The USKF is a modified version of the SKF, adapted for nonlinear systems, where the bias is not exactly estimated rather its correlations with the state are updated. The CUKF is based on two major functional components. First, it resorts to the UKF for estimation. Subsequently, it draws sigma points based on these estimates which are projected onto a region, by solving an optimization problem, supposing a constraint that the measurements can only be positively biased. In our numerical evaluation, we assume perfect detection for these two algorithms. Note that in the implementation of the USKF and CUKF, we switch to the standard UKF when no bias is detected. Also note that the proposed BDM filter has no limitations in terms of whether any bias positively or negatively disturbs the measurements. However, since the CUKF assumes a positive bias we keep this restriction in our simulations. In particular, bias \textcolor{black}{in each dimension }is added as a shifted Gaussian $\mathcal{N}(\mu,\sigma^2)$ where $\mu \geq 0$ \cite{filtermobile}.

\textcolor{black}{For performance evaluation, we resort to a target tracking problem in a wireless network where the range measurements are typically biased \cite{weiss2008network}. The observations get biased owing to the transitions between line-of-sight (LOS) and non-line-of-sight (NLOS) conditions \cite{4027766}. Such tracking systems find applications in emergency services, fleet management, intelligent transportation, etc. \cite{4960267}.}
	
\textcolor{black}{We consider the process equation for the target assuming an unknown turning rate as \cite{8398426}}

\begin{flalign}
	\mathbf{x}_k=\mathbf{f}(\mathbf{x}_{k-1})+\mathbf{q}_{k-1}\label{eqn_res1a}
\end{flalign}
with
\begin{align}
	\mathbf{f}(\mathbf{x}_{k-1}) &= \begin{pmatrix} \text{1} & \frac{\text{sin}(\omega_{k}{\textcolor{black}{\zeta_t}})}{\omega_{k}} & \text{0} &  \frac{\text{cos}(\omega_{k}{\textcolor{black}{\zeta_t}})-\text{1}}{\omega_{k}} & \text{0} \\  \text{0} & \text{cos}(\omega_{k}{\textcolor{black}{\zeta_t}}) &  \text{0} & -\text{sin}(\omega_{k}{\textcolor{black}{\zeta_t}}) & \text{0}\\  \text{0} & \frac{\text{1}-\text{cos}(\omega_{k}{\textcolor{black}{\zeta_t}})}{\omega_{k}} & \text{1} &\frac{\text{sin}(\omega_{k}{\textcolor{black}{\zeta_t}})}{\omega_{k}} & \text{0} \\  \text{0} &  \text{\text{sin}}(\omega_{k}{\textcolor{black}{\zeta_t}}) &  \text{0} &  \text{cos}(\omega_{k}{\textcolor{black}{\zeta_t}}) & \text{0}\\  \text{0} & \text{0} & \text{0} & \text{0} &\text{1} \end{pmatrix} \mathbf{x}_{k-\text{1}}\label{eqn_res1}
\end{align}
where the state vector $\mathbf{x}_k= [a_k,\dot{{a_k}},b_k,\dot{{b_k}},\omega_{k}]^{\top}$
contains the 2D position coordinates \(({a_k} , {b_k} )\), the respective velocities \((\dot{{a_k}} , \dot{{b_k}} )\), the angular velocity $\omega_{k}$ of the target at time instant $k$, \( {\textcolor{black}{\zeta_t}} \) is the sampling period, and $\mathbf{q}_{k-\text{1}} \sim N\left(0,\mathbf{Q}_{k-\text{1}}.\right)$. $\mathbf{Q}_{k-\text{1}}$ is given in terms of scaling parameters $\eta_1$ and $\eta_2$ as \cite{8398426}
\begin{equation}
	\mathbf{Q}_{k-\text{1}}=\begin{pmatrix} \eta_1 \mathbf{M} & 0 & 0\\0 &\eta_1 \mathbf{M}&0\\0&0&\eta_2
	\end{pmatrix}, \mathbf{M}=\begin{pmatrix} {\textcolor{black}{\zeta_t}}^3/3 & {\textcolor{black}{\zeta_t}}^2/2\\{\textcolor{black}{\zeta_t}}^2/2 &{\textcolor{black}{\zeta_t}}
	\end{pmatrix}\nonumber
\end{equation}

Range readings are obtained from $m$ sensors installed around a rectangular area where the $i$th sensor is located at $\big(a^{\rho_i}=350(i-1),b^{\rho_i}=350\ ((i-1)\mod2)\big)$. The nominal measurement equation can therefore be expressed as
\begin{equation}
	\mathbf{y}_k = \mathbf{h}(\mathbf{x}_{k})+\mathbf{r}_k\label{eqn_res2}
\end{equation}
with
\begin{equation}
		{h^i(\mathbf{x}_k )} = \sqrt{\big( (a_{k} - a^{\rho_i})^{2} + (b_{k} - b^{\rho_i})^{2} \big)}\label{eqn_res2b}
\end{equation}
{For the duration of bias presence the following observation equations, based on the random bias model \cite{park2022robust}, are assumed} 
\begin{align}
	\mathbf{y}_k &= \mathbf{h}(\mathbf{x}_{k})+\mathbf{r}_k+ \overbrace{\boldsymbol{{\mathcal{J}}}_{k} (\mathbf{o}_k+\triangle \mathbf{o}_k)}^{\mathbf{b}_k}\label{eqn_res2c}\\
	\mathbf{y}_{k+1} &= \mathbf{h}(\mathbf{x}_{k+1})+\mathbf{r}_{k+1}+ \boldsymbol{{\mathcal{J}}}_{k} (\mathbf{o}_k+\triangle \mathbf{o}_{k+1})\label{eqn_res2d}
\end{align}
{where $\mathbf{o}_k\in\mathbb{R}^m$ models the effect of statistically independent biases in the measurements and $\boldsymbol{\mathcal{J}}_{k} \in \mathbb{R}^{m\times m}$ is a diagonal matrix with independent Bernoulli elements with parameter $\lambda$.} \textcolor{black}{Our measurement model for evaluation is inspired by the LOS and NLOS models in \cite{4027766,4960267} where shifted Gaussian distributions are used to model the biased measurements. In effect, bias with magnitude ${o}^i_k$ with some additional uncertainty $\triangle o^i_k \sim \mathcal{N}({0},\Sigma_\mathbf{o}^i)$ affects the \textit{i}th dimension at time step $k$ if ${\mathcal{J}}^i_{k}=1$. Since the exact magnitudes of biases are not generally known apriori and can occur randomly in a given range we assume $o^i_k$ to be uniformly distributed i.e. ${o}^i_k\sim\mathcal{U}(0,\Lambda^i)$. This is in contrast to the approaches in \cite{4027766,4960267} where the magnitudes of biases are assumed to be known perfectly. Similarly at time step $k+1$, the bias sustains with the same magnitude of the previous time step $k$ in the \textit{i}th dimension i.e. ${o}^i_k$ with some uncertainty $\triangle o^i_{k+1}$ if the \textit{i}th dimension at the previous time step $k$ was affected i.e. ${\mathcal{J}}^i_{k}=1$. Note that the evaluation model allows us to compare different methods under extreme conditions varying from the case of no biased dimension to the case where every observation can possibly get biased.}


For evaluation we assign the following values to different parameters: ${\Sigma_\mathbf{o}^i}  = {0.4}, \Lambda^i=90, \mathbf{x}_{0} = [0, 10, 0, -5, \frac{3 \pi}{180}]^{\top}, \mathbf{R}_k = 4\mathbf{I},  {\textcolor{black}{\zeta_t}}  = 1, \eta_1=0.1, \eta_2=1.75\times10^{-4}$. We assume $m=4$, since for higher-dimensional problems even rejection-based methods like the SOR-UKF can have acceptable performance for a larger probability of errors owing to the redundancy of useful information in other uncorrupted dimensions. However, this does not limit the applicability of the proposed method for large $m$. Our point is to emphasize how properly utilizing information from an affected dimension, a characteristic of {analytical} redundancy approaches, is more useful in contrast to completely rejecting the information, a characteristic of {hardware} redundancy approaches \cite{chughtai2020robust}. In addition, we evaluate the relative performance of the proposed filter with similar {analytical} redundancy approaches. 

\begin{figure*}[ht!]
	\centering
	\begin{subfigure}[h!]{0.22\textwidth}
		\centering
		\includegraphics[width=\linewidth,trim=0 0 20cm 0,clip=true]{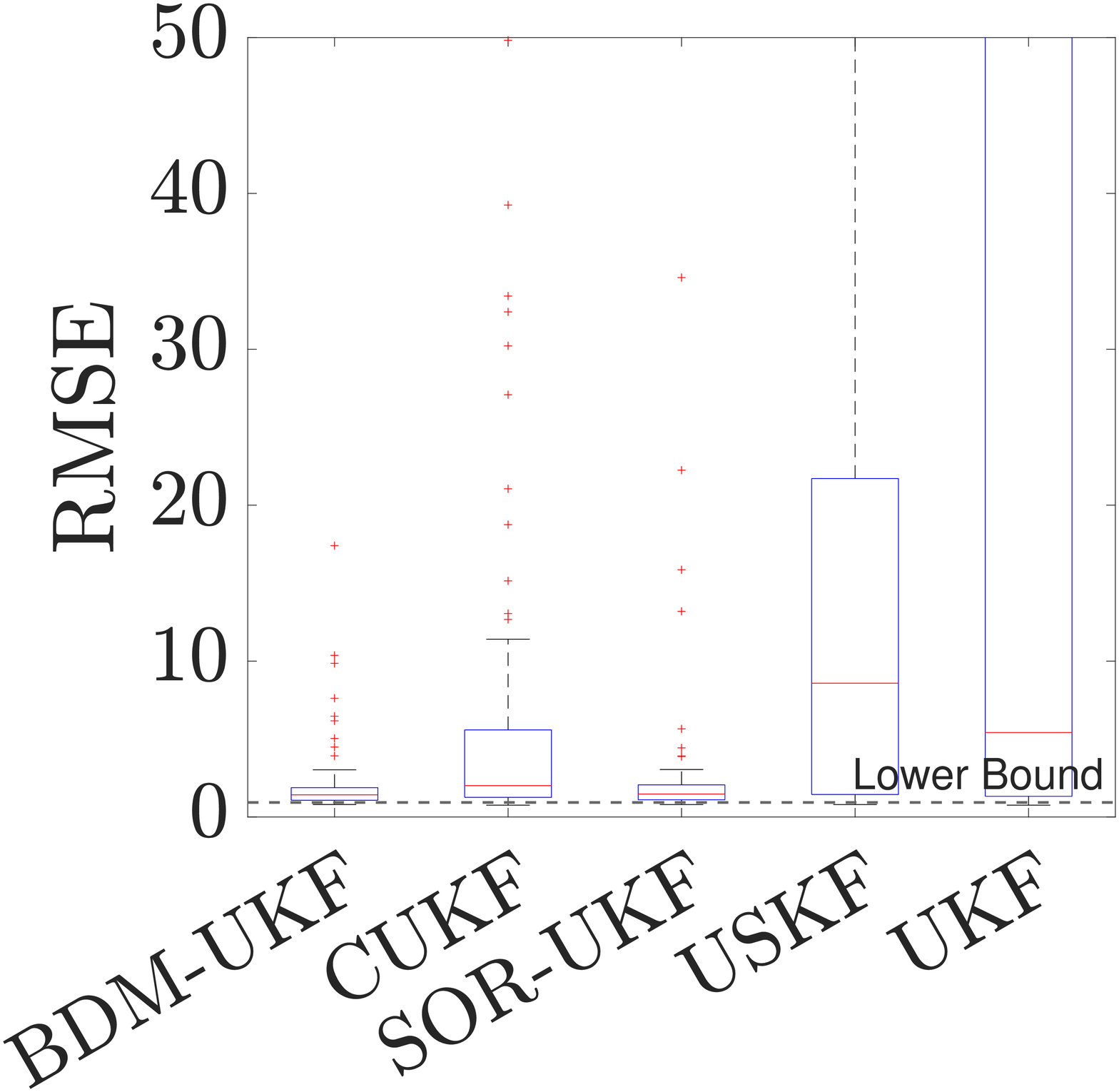}
		\caption{$\lambda = 0.2$}
		\label{Box11}
	\end{subfigure}
	\begin{subfigure}[h!]{0.22\textwidth}
		\centering
		\includegraphics[width=\linewidth,trim=0 0 20cm 0,clip=true]{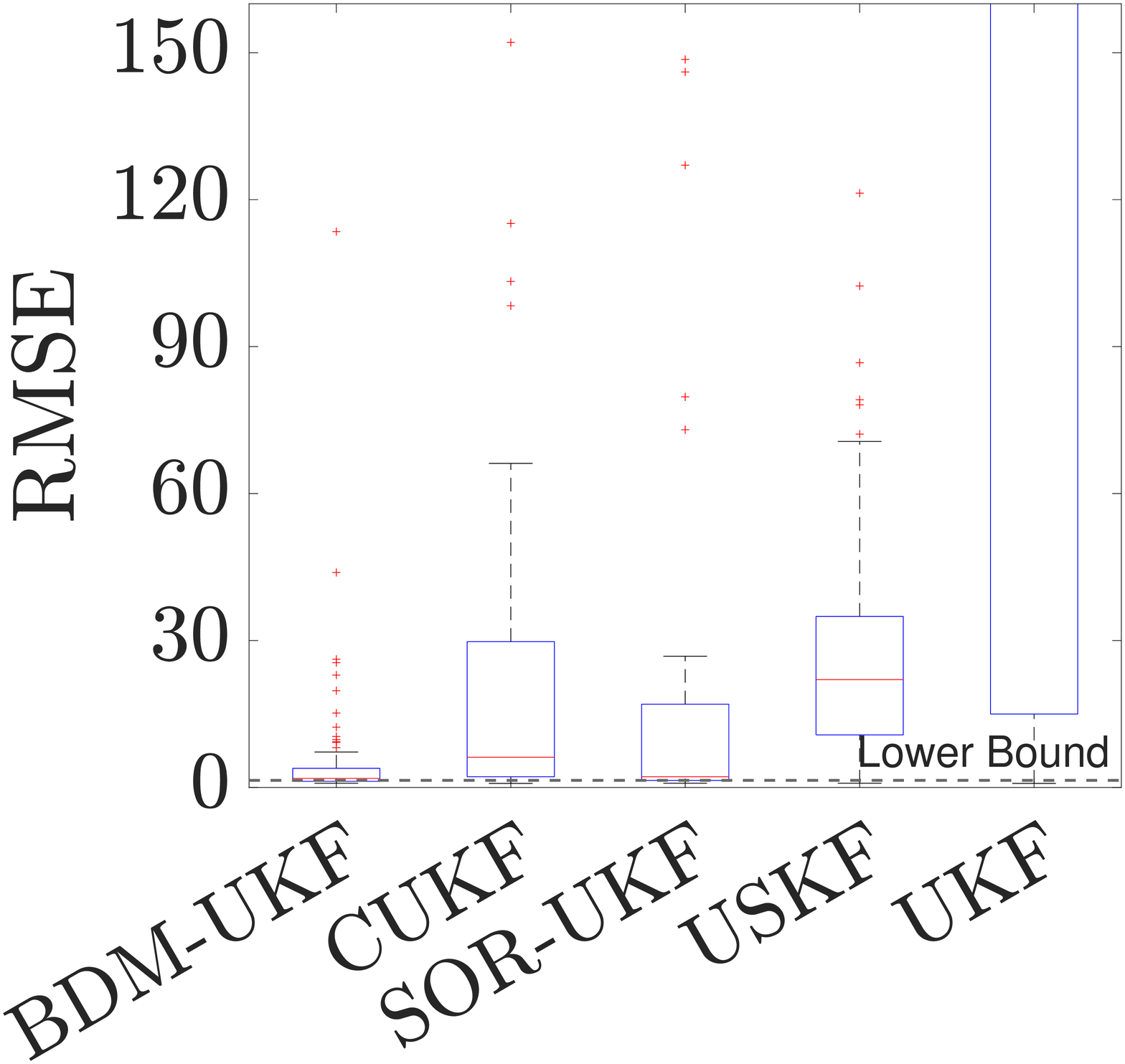}
		\caption{$\lambda = 0.4$}
		\label{Box12}
	\end{subfigure}
	\begin{subfigure}[h!]{0.22\textwidth}
		\centering
		\includegraphics[width=\linewidth,trim=0 0 20cm 0,clip=true]{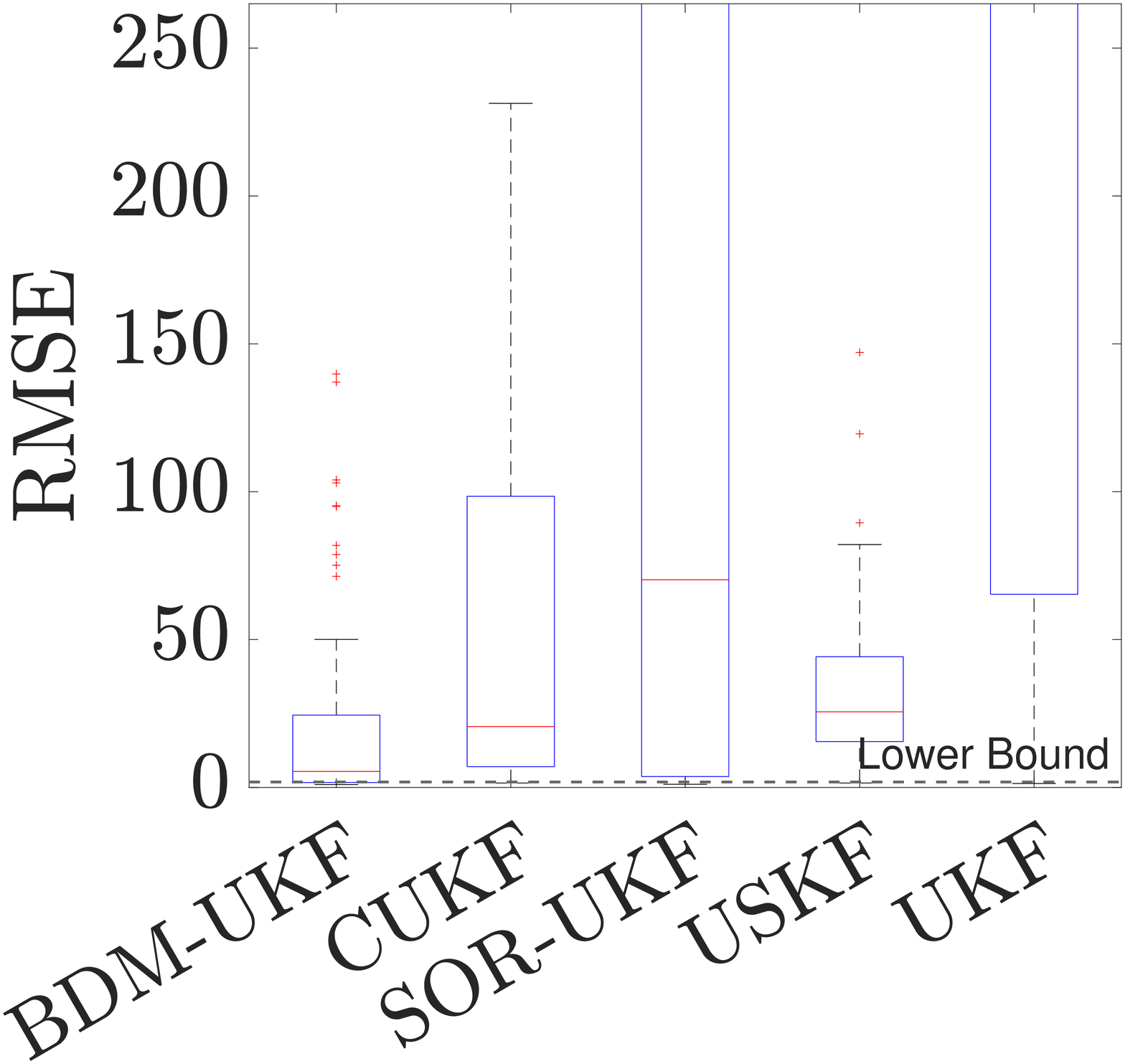}
		\caption{$\lambda = 0.6$}
		\label{Box13}
	\end{subfigure}
	\begin{subfigure}[h!]{0.22\textwidth}
		\centering
		\includegraphics[width=\linewidth,trim=0 0 20cm 0,clip=true]{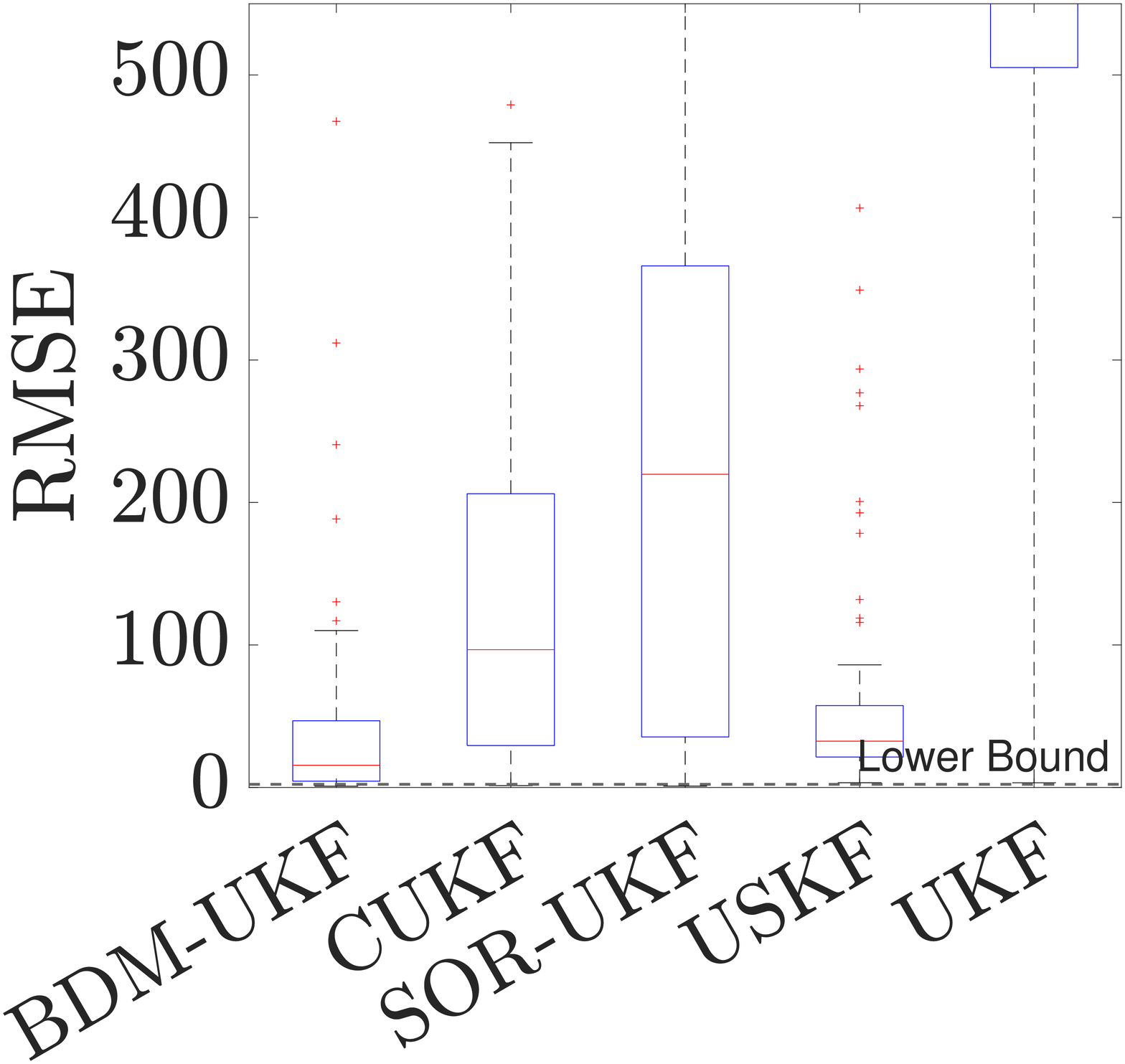}
		\caption{$\lambda = 0.8$}
		\label{Box14}
	\end{subfigure}
	\caption{Box plots of state RMSE for Case 1 with increasing $\lambda$} \label{Box1}
\end{figure*}
\begin{figure}[ht!]
	\centering
	\includegraphics[width=\linewidth]{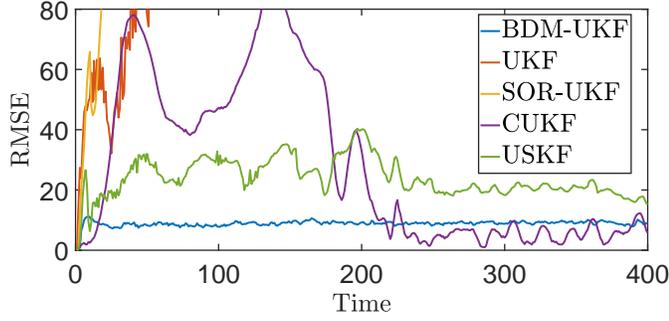}
	\caption{Tracking performance of algorithms over time for an example MC run for Case 1} 
	\label{fig:Case1_run}
\end{figure}

For each method, the UKF parameters are set as $\alpha=1$, $\beta=2$ and $\kappa=0$. We consider $T = 400$ time steps and carry out 100 Monte Carlo (MC) simulations for each case. All the filters are initialized with a state estimate equal to $\mathbf{x}_{0}$ its error covariance as $\mathbf{Q}_{k}$ and the bias estimate is set equal to zero initially. The other parameters supposed for the BDM-UKF are: $\mathbf{\Sigma}_{0}^{+}=0.001\mathbf{I},\widetilde{\mathbf{\Sigma}}_{k}=1000\mathbf{R}_k,\breve{\mathbf{\Sigma}}_{k}=0.1\mathbf{R}_k$ and $\theta^i_k=0.5~\forall~i$ and $\tau=10^{-4}$. In CUKF the auxiliary parameters are assigned values as: $\epsilon=1,\sigma_n=2$. Lastly, the USKF is initialized with a zero mean bias with a covariance of $500\mathbf{I}$ having zero cross-covariance with the state. Other parameters for all the filters are kept the same as originally reported. 
\subsection*{Performance bounds}
To benchmark the relative performance of the considered algorithms, we evaluate the PCRB for the target tracking model in \eqref{eqn_res1a}-\eqref{eqn_res2c}. The bound remains valid where system dynamics can be modeled as \eqref{eqn_res1a} and where the measurement model can be considered to switch between \eqref{eqn_res2} and \eqref{eqn_res2c}-\eqref{eqn_res2d} since we use a generalized model in our derivation. We evaluate the PCRB assuming perfect apriori knowledge of the occurrence time, duration of the bias and $\lambda$, resorting to the approach presented in \cite{668800,van2013detection} for a generalized nonlinear dynamical system, with white process and measurement noise, given as follows
\begin{align}
        \mathbf{x}_{k+1}&=\mathbf{f}\left(\mathbf{x}_{k}, \mathbf{q}_{k}\right)\\
		\mathbf{y}_{k}&=\mathbf{h}\left(\mathbf{x}_{k}, \mathbf{r}_{k},\mathbf{y}_{k-1},\cdots,\mathbf{y}_{k-z}\right)
\end{align} 
where $z$ is a positive integer.		
The PCRB matrix for the estimation error of $\mathbf{x}_k$ can be written as 
\begin{equation}
	\text{PCRB}_k\triangleq{\mathbf{J}_k}^{-1}\label{PCRB_1}
\end{equation}	
where $\mathbf{J}_k$ can be expressed recursively as 
\begin{align}
\mathbf{J}_{k+1}&=\mathbf{D}_{k}^{22}-\mathbf{D}_{k}^{21}\left(\mathbf{J}_{k}+\mathbf{D}_{k}^{11}\right)^{-1} \mathbf{D}_{k}^{12}\label{PCRB_2}
\end{align}
with
\begin{align}
&\mathbf{D}_{k}^{11}=\langle-\Delta_{\mathbf{x}_{k}}^{\mathbf{x}_{k}} \log p\left(\mathbf{x}_{k+1} \mid \mathbf{x}_{k}\right)\rangle_{p(\mathbf{x}_{k+1},\mathbf{x}_{k})}\label{PCRB_3}\\
&\mathbf{D}_{k}^{12}=\langle-\Delta_{\mathbf{x}_{k}}^{\mathbf{x}_{k+1}} \log p\left(\mathbf{x}_{k+1} \mid \mathbf{x}_{k}\right)\rangle_{p(\mathbf{x}_{k+1},\mathbf{x}_{k})}\label{PCRB_4}\\
&\mathbf{D}_{k}^{21}=\langle-\Delta_{\mathbf{x}_{k+1}}^{\mathbf{x}_{k}} \log p\left(\mathbf{x}_{k+1} \mid \mathbf{x}_{k}\right)\rangle_{p(\mathbf{x}_{k+1},\mathbf{x}_{k})}=\left[\mathbf{D}_{k}^{12}\right]^{\top}\label{PCRB_5}\\
&\mathbf{D}_{k}^{22}= \mathbf{D}_{k}^{22}(1)+\mathbf{D}_{k}^{22}(2)\\
&\mathbf{D}_{k}^{22}(1)=\langle-\Delta_{\mathbf{x}_{k+1}}^{\mathbf{x}_{k+1}} \log p\left(\mathbf{x}_{k+1} \mid \mathbf{x}_{k}\right)\rangle_{p(\mathbf{x}_{k+1},\mathbf{x}_{k})} \nonumber\\
&\mathbf{D}_{k}^{22}(2)=\nonumber\\
&\langle-\Delta_{\mathbf{x}_{k+1}}^{\mathbf{x}_{k+1}} \log p\left(\mathbf{y}_{k+1} \mid \mathbf{x}_{k+1},\mathbf{y}_{k},\cdots,\mathbf{y}_{k-z+1}\right)\rangle_{p(\mathbf{y}_{k+1},\mathbf{x}_{k+1},\cdots)} \label{PCRB_6}
\end{align}
where
\begin{align}
\Delta_{\Psi}^{\Theta}&=\nabla_{\Psi} \nabla_{\Theta}^{\top}	\label{PCRB_8}\\
\nabla_{\Theta}&=\left[\frac{\partial}{\partial \Theta_{1}}, \cdots, \frac{\partial}{\partial \Theta_{r}}\right]^{\top}\label{PCRB_9}
\end{align}
The bound is valid given the existence of derivatives and expectations terms in \eqref{PCRB_1}-\eqref{PCRB_9} for an asymptotically unbiased estimator \cite{668800}. Using results from \cite{668800,van2013detection} we obtain 
\begin{align}
	&\mathbf{D}_{k}^{11}=\left[\nabla_{\mathbf{x}_{k}} \mathbf{f}^{\top}\left(\mathbf{x}_{k}\right)\right] \mathbf{Q}_{k}^{-1}\left[\nabla_{\mathbf{x}_{k}} \mathbf{f}^{\top}\left(\mathbf{x}_{k}\right)\right]^{\top}\label{PCRB_10}\\
	&\mathbf{D}_{k}^{12}=-\nabla_{\mathbf{x}_{k}} \mathbf{f}^{\top}\left(\mathbf{x}_{k}\right) \mathbf{Q}_{k}^{-1}\label{PCRB_11}\\
	&\mathbf{D}_{k}^{22}(1)=\mathbf{Q}_{k}^{-1}
\end{align}

For the period where the nominal equation \eqref{eqn_res2} remains applicable, $\mathbf{D}_{k}^{22}(2)$ can be expressed as \cite{van2013detection}
\begin{align}
	&\mathbf{D}_{k}^{22}(2)=\nonumber
	\\&-\mathbf{Q}_{k}^{-1} \langle \tilde{\mathbf{F}}_{k} \rangle_{p(\mathbf{x}_{k})}  \left[\mathbf{J}_{k}+ \langle \tilde{\mathbf{F}}_{k}^{\top} \mathbf{Q}_{k}^{-1} \tilde{\mathbf{F}}_{k} \rangle_{p(\mathbf{x}_{k})} \right]^{-1} \langle \tilde{\mathbf{F}}_{k}^{\top}\rangle_{p(\mathbf{x}_{k})}\mathbf{Q}_{k}^{-1} \nonumber \\
	&+\langle \tilde{\mathbf{H}}_{k+1}^{\top} \mathbf{R}_{k+1}^{-1} \tilde{\mathbf{H}}_{k+1}\rangle_{p(\mathbf{x}_{k+1})}
\end{align}
where $\tilde{\mathbf{F}}_{k}$ and $\tilde{\mathbf{H}}_{k}$ are the Jacobians of $\mathbf{f}(\mathbf{x}_{k})$ and $\mathbf{h}(\mathbf{x}_{k})$ respectively. 

For the duration of bias occurrence, $\mathbf{D}_{k}^{22}(2)$ can be evaluated using \eqref{PCRB_6}, equivalently written as follows using a technique similar to \cite{6916255}
\begin{equation}
\mathbf{D}_{k}^{22}(2)=\big\langle \frac{[\nabla_{\mathbf{x}_{k+1}} p\left(\mathbf{y}_{k+1} \mid \mathbf{x}_{k+1},\mathbf{y}_{k},\mathbf{x}_{k}\right)][.]^\top}{[p\left(\mathbf{y}_{k+1} \mid \mathbf{x}_{k+1},\mathbf{y}_{k},\mathbf{x}_{k}\right)]^2}	\big\rangle_{p(\mathbf{y}_{k+1},\mathbf{x}_{k+1},\cdots)}
\end{equation}

First consider \eqref{eqn_res2c} to evaluate $\mathbf{D}_{k}^{22}(2)$ at the instance of bias occurrence for which $p\left(\mathbf{y}_{k+1} \mid \mathbf{x}_{k+1},\mathbf{y}_{k},\mathbf{x}_{k}\right) $ can be approximated using Monte Carlo method as
\begin{align}
	&p\left(\mathbf{y}_{k+1} \mid \mathbf{x}_{k+1},\mathbf{y}_{k},\mathbf{x}_{k}\right)=	p\left(\mathbf{y}_{k+1} \mid \mathbf{x}_{k+1}\right) \\ 
	&=\int p\left(\mathbf{y}_{k+1} \mid \mathbf{x}_{k+1},\mathbf{b}_{k+1}\right)p(\mathbf{b}_{k+1})d\mathbf{b}_{k+1}\\
	&=\int \mathcal{N}\left(\mathbf{y}_{k+1}|\mathbf{h}(\mathbf{x}_{k+1})+\mathbf{b}_{k+1},\mathbf{R}_k\right) p(\mathbf{b}_{k+1})d\mathbf{b}_{k+1}\\
	&\approx \frac{1}{N_{mc1}}\sum_i\mathcal{N}\left(\mathbf{y}_{k+1}|\mathbf{h}(\mathbf{x}_{k+1})+\mathbf{b}^{(i)}_{k+1},\mathbf{R}_k\right) 
\end{align}
where $\mathbf{b}^{(i)}_{k+1}$, $i=1,\cdots,N_{mc1}$,  are independent and identically distributed (i.i.d.) samples such that $\mathbf{b}^{(i)}_{k+1}\sim p(\mathbf{b}_{k+1})$.

Accordingly, $\mathbf{D}_{k}^{22}(2)$ can be approximated as
\begin{equation}
	\mathbf{D}_{k}^{22}(2)=\big\langle \frac{[\nabla_{\mathbf{x}_{k+1}} \sum_i \exp(\phi^{(i)})][.]^\top}{[\sum_i \exp(\phi^{(i)})]^2}	\big\rangle_{p(\mathbf{y}_{k+1},\mathbf{x}_{k+1})}
\end{equation}
where $\phi^{(i)}=-0.5(\mathbf{y}_{k+1}-(\mathbf{h}(\mathbf{x}_{k+1})+\mathbf{b}^{(i)}_{k+1}))^{\top}\mathbf{R}_{k+1}^{-1}(\mathbf{y}_{k+1}-(\mathbf{h}(\mathbf{x}_{k+1})+\mathbf{b}^{(i)}_{k+1}))$. We can further write
\begin{equation}
	\mathbf{D}_{k}^{22}(2)=\big\langle \frac{\sum_i [\exp(\phi^{(i)})\nabla_{\mathbf{x}_{k+1}}\phi^{(i)}  ][.]^\top}{[\sum_i \exp(\phi^{(i)})]^2}	\big\rangle_{p(\mathbf{y}_{k+1},\mathbf{x}_{k+1})}
\end{equation}
where $\nabla_{\mathbf{x}_{k+1}}\phi^{(i)}=\tilde{\mathbf{H}}_{k+1}^{\top}\mathbf{R}_{k+1}^{-1}(\mathbf{y}_{k+1}-\mathbf{h}(\mathbf{x}_{k+1})-\mathbf{b}^{(i)}_{k+1})$

Resultingly, $\mathbf{D}_{k}^{22}(2)$ is approximated as
\begin{equation}
	\mathbf{D}_{k}^{22}(2)\approx\frac{1}{N_{mc2}}\sum_j \frac{\sum_i [\exp(\phi^{(i,j)})\nabla_{\mathbf{x}_{k+1}}\phi^{(i,j)} ][.]^\top}{[\sum_i \exp(\phi^{(i,j)})]^2}
\end{equation}
with $\phi^{(i,j)}=-0.5(\mathbf{y}^{(j)}_{k+1}-(\mathbf{h}(\mathbf{x}^{(j)}_{k+1})+\mathbf{b}^{(i)}_{k+1}))^{\top}\mathbf{R}_{k+1}^{-1}(\mathbf{y}^{(j)}_{k+1}-(\mathbf{h}(\mathbf{x}^{(j)}_{k+1})+\mathbf{b}^{(i)}_{k+1}))$ and $\nabla_{\mathbf{x}_{k+1}}\phi^{(i,j)}=\tilde{\mathbf{H}}_{k+1}^{\top}\mathbf{R}_{k+1}^{-1}(\mathbf{y}^{(j)}_{k+1}-\mathbf{h}(\mathbf{x}^{(j)}_{k+1})-\mathbf{b}^{(i)}_{k+1})$. $\mathbf{y}^{(j)}_{k+1},\mathbf{x}^{(j)}_{k+1}$, $j=1,\cdots,N_{mc2}$, are i.i.d. samples such that $(\mathbf{y}^{(j)}_{k+1},\mathbf{x}^{(j)}_{k+1})\sim p(\mathbf{y}_{k+1},\mathbf{x}_{k+1})$.

Lastly, to evaluate $\mathbf{D}_{k}^{22}(2)$ for the bias persistence period we consider \eqref{eqn_res2d} and the difference of \eqref{eqn_res2c}-\eqref{eqn_res2d}.
Therefore, $p\left(\mathbf{y}_{k+1} \mid \mathbf{x}_{k+1},\mathbf{y}_{k},\mathbf{x}_{k}\right) $ can be approximated for this case using Monte Carlo method as
\begin{align}
	&p\left(\mathbf{y}_{k+1} \mid \mathbf{x}_{k+1},\mathbf{y}_{k},\mathbf{x}_{k}\right)\\ 
	&=\int p\left(\mathbf{y}_{k+1} \mid \mathbf{x}_{k+1},\mathbf{y}_{k},\mathbf{x}_{k},\boldsymbol{{\mathcal{J}}}_{k}\right)p(\boldsymbol{{\mathcal{J}}}_{k})d\boldsymbol{{\mathcal{J}}}_{k}\\
	&\approx \frac{1}{N_{mc3}}\sum_i\mathcal{N}(\mathbf{y}_{k+1}|\mathbf{h}(\mathbf{x}_{k+1})+\boldsymbol{{\mathcal{J}}}^{(i)}_{k}(\mathbf{y}_k-\mathbf{h}(\mathbf{x}_{k}))\nonumber \\ 
	&\ \ \ \ \ \ \ \ \ \ \ \ \ \ \ \ \ \ \ \ , \mathbf{R}_{k+1}+\boldsymbol{{\mathcal{J}}}^{(i)}_{k}(\mathbf{R}_{k}+2\Sigma_\mathbf{o})) 
\end{align}
where $\boldsymbol{{\mathcal{J}}}^{(i)}_{k}$, $i=1,\cdots,N_{mc3}$,  are i.i.d. samples such that $\boldsymbol{{\mathcal{J}}}^{(i)}_{k}\sim p(\boldsymbol{{\mathcal{J}}}_{k})$. Resultingly, $\mathbf{D}_{k}^{22}(2)$ is approximated as
\begin{equation}
	\mathbf{D}_{k}^{22}(2)=\big\langle \frac{[\nabla_{\mathbf{x}_{k+1}} \sum_i \exp(\theta^{(i)})][.]^\top}{[\sum_i \exp(\theta^{(i)})]^2}	\big\rangle_{p(\mathbf{y}_{k+1},\mathbf{x}_{k+1},\cdots)}
\end{equation}
where $\theta^{(i)}=-0.5(\mathbf{y}_{k+1}-(\mathbf{h}(\mathbf{x}_{k+1})+\boldsymbol{{\mathcal{J}}}^{(i)}_{k}(\mathbf{y}_k-\mathbf{h}(\mathbf{x}_{k})))^{\top}{(\mathbf{R}_{k+1}+\boldsymbol{{\mathcal{J}}}^{(i)}_{k}(\mathbf{R}_{k}+2\Sigma_\mathbf{o}))}^{-1}(\mathbf{y}_{k+1}-(\mathbf{h}(\mathbf{x}_{k+1})+\boldsymbol{{\mathcal{J}}}^{(i)}_{k}(\mathbf{y}_k-\mathbf{h}(\mathbf{x}_{k})))$.
Furthermore
\begin{equation}
	\mathbf{D}_{k}^{22}(2)=\big\langle \frac{\sum_i [\exp(\theta^{(i)})\nabla_{\mathbf{x}_{k+1}}\theta^{(i)}  ][.]^\top}{[\sum_i \exp(\theta^{(i)})]^2}	\big\rangle_{p(\mathbf{y}_{k+1},\mathbf{x}_{k+1},\cdots)}
\end{equation}
where $\nabla_{\mathbf{x}_{k+1}}\theta^{(i)}=\tilde{\mathbf{H}}_{k+1}^{\top}{(\mathbf{R}_{k+1}+\boldsymbol{{\mathcal{J}}}^{(i)}_{k}(\mathbf{R}_{k}+2\Sigma_\mathbf{o}))}^{-1}(\mathbf{y}_{k+1}-(\mathbf{h}(\mathbf{x}_{k+1})+\boldsymbol{{\mathcal{J}}}^{(i)}_{k}(\mathbf{y}_k -\mathbf{h}(\mathbf{x}_{k}) )))$. Resultingly, $\mathbf{D}_{k}^{22}(2)$ is approximated as
\begin{equation}
	\mathbf{D}_{k}^{22}(2)\approx\frac{1}{N_{mc4}}\sum_j \frac{\sum_i [\exp(\theta^{(i,j)})\nabla_{\mathbf{x}_{k+1}}\theta^{(i,j)} ][.]^\top}{[\sum_i \exp(\theta^{(i,j)})]^2}
\end{equation}
with $\theta^{(i,j)}=-0.5(\mathbf{y}^{(j)}_{k+1}-(\mathbf{h}(\mathbf{x}^{(j)}_{k+1})+\boldsymbol{{\mathcal{J}}}^{(i)}_{k}(\mathbf{y}^{(j)}_k-\mathbf{h}(\mathbf{x}^{(j)}_{k})))^{\top}{(\mathbf{R}_{k+1}+\boldsymbol{{\mathcal{J}}}^{(i)}_{k}(\mathbf{R}_{k}+2\Sigma_\mathbf{o}))}^{-1}(\mathbf{y}^{(j)}_{k+1}-(\mathbf{h}(\mathbf{x}^{(j)}_{k+1})+\boldsymbol{{\mathcal{J}}}^{(i)}_{k}(\mathbf{y}^{(j)}_k-\mathbf{h}(\mathbf{x}^{(j)}_{k})))$ and
$\nabla_{\mathbf{x}_{k+1}}\theta^{(i,j)}=\tilde{\mathbf{H}}_{k+1}^{\top}{(\mathbf{R}_{k+1}+\boldsymbol{{\mathcal{J}}}^{(i)}_{k}(\mathbf{R}_{k}+2\Sigma_\mathbf{o}))}^{-1}(\mathbf{y}^{(j)}_{k+1}-(\mathbf{h}(\mathbf{x}^{(j)}_{k+1})+\boldsymbol{{\mathcal{J}}}^{(i)}_{k}(\mathbf{y}^{(j)}_k -\mathbf{h}(\mathbf{x}^{(j)}_{k}) )))$.
$\mathbf{y}^{(j)}_{k+1},\mathbf{x}^{(j)}_{k+1},\mathbf{y}^{(j)}_{k},\mathbf{x}^{(j)}_{k}$, $j=1,\cdots,N_{mc4}$, are i.i.d. samples such that $(\mathbf{y}^{(j)}_{k+1},\mathbf{x}^{(j)}_{k+1},\mathbf{y}^{(j)}_{k},\mathbf{x}^{(j)}_{k})\sim p(\mathbf{y}_{k+1},\mathbf{x}_{k+1},\mathbf{y}_{k},\mathbf{x}_{k})$.

\begin{figure*}[t!]
	\centering
	\begin{subfigure}[t!]{0.22\textwidth}
		\centering
		\includegraphics[width=\linewidth,trim=0 0 20cm 0,clip=true]{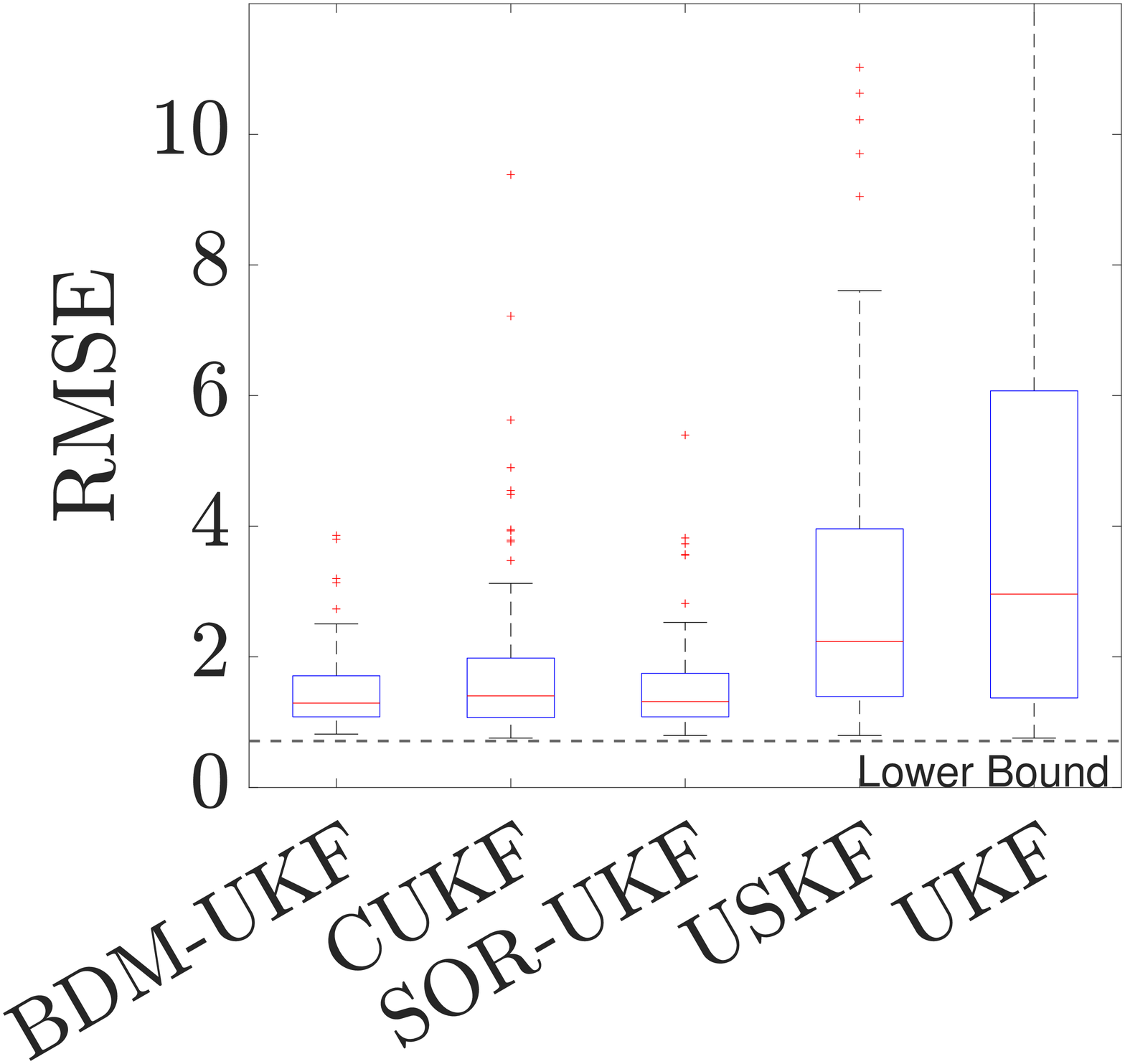}
		\caption{$\lambda = 0.2$}
		\label{Box21}
	\end{subfigure}
	\begin{subfigure}[t!]{0.22\textwidth}
		\centering
		\includegraphics[width=\linewidth,trim=0 0 20cm 0,clip=true]{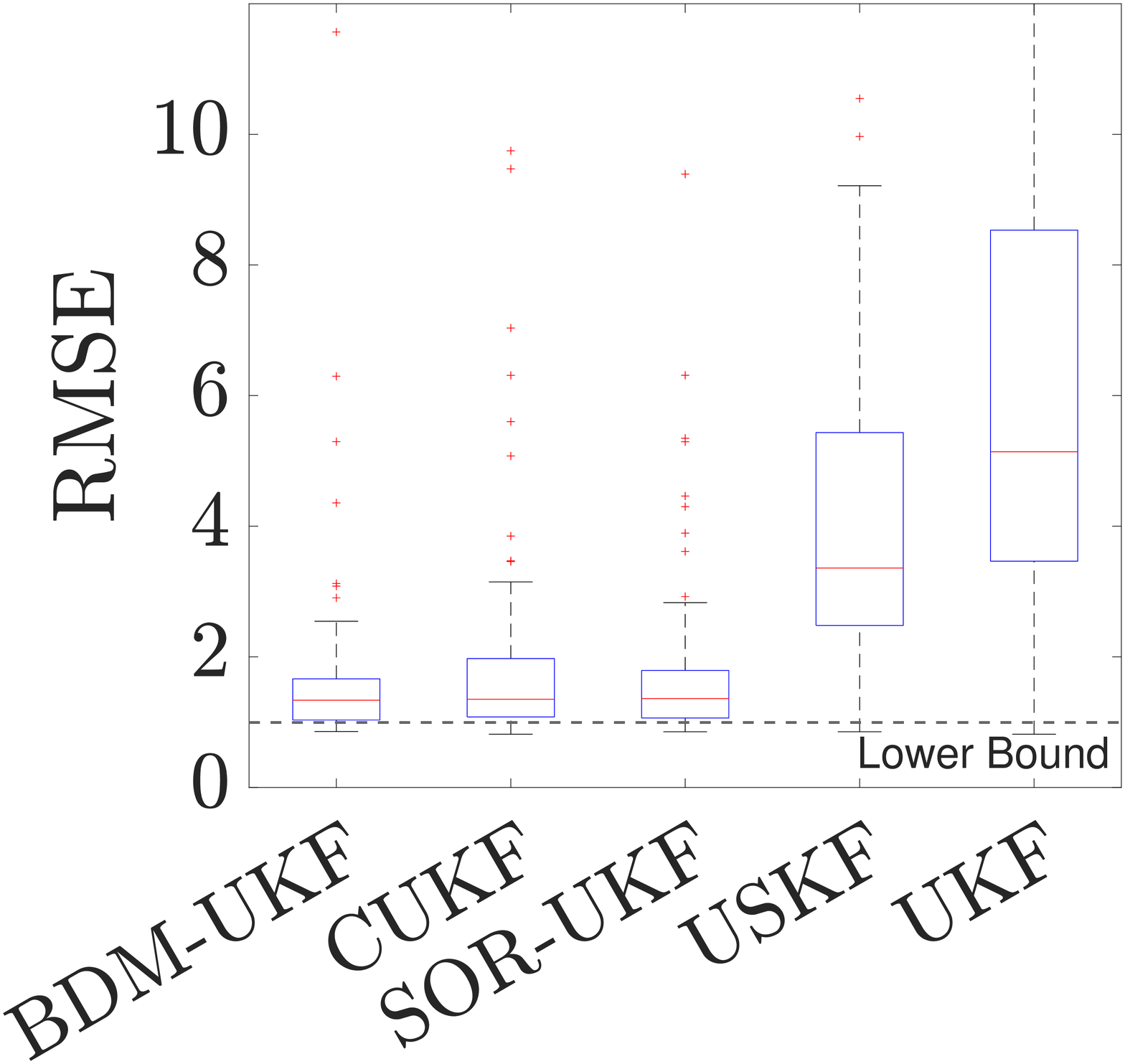}
		\caption{$\lambda = 0.4$}
		\label{Box22}
	\end{subfigure}
	\begin{subfigure}[t!]{0.22\textwidth}
		\centering
		\includegraphics[width=\linewidth,trim=0 0 20cm 0,clip=true]{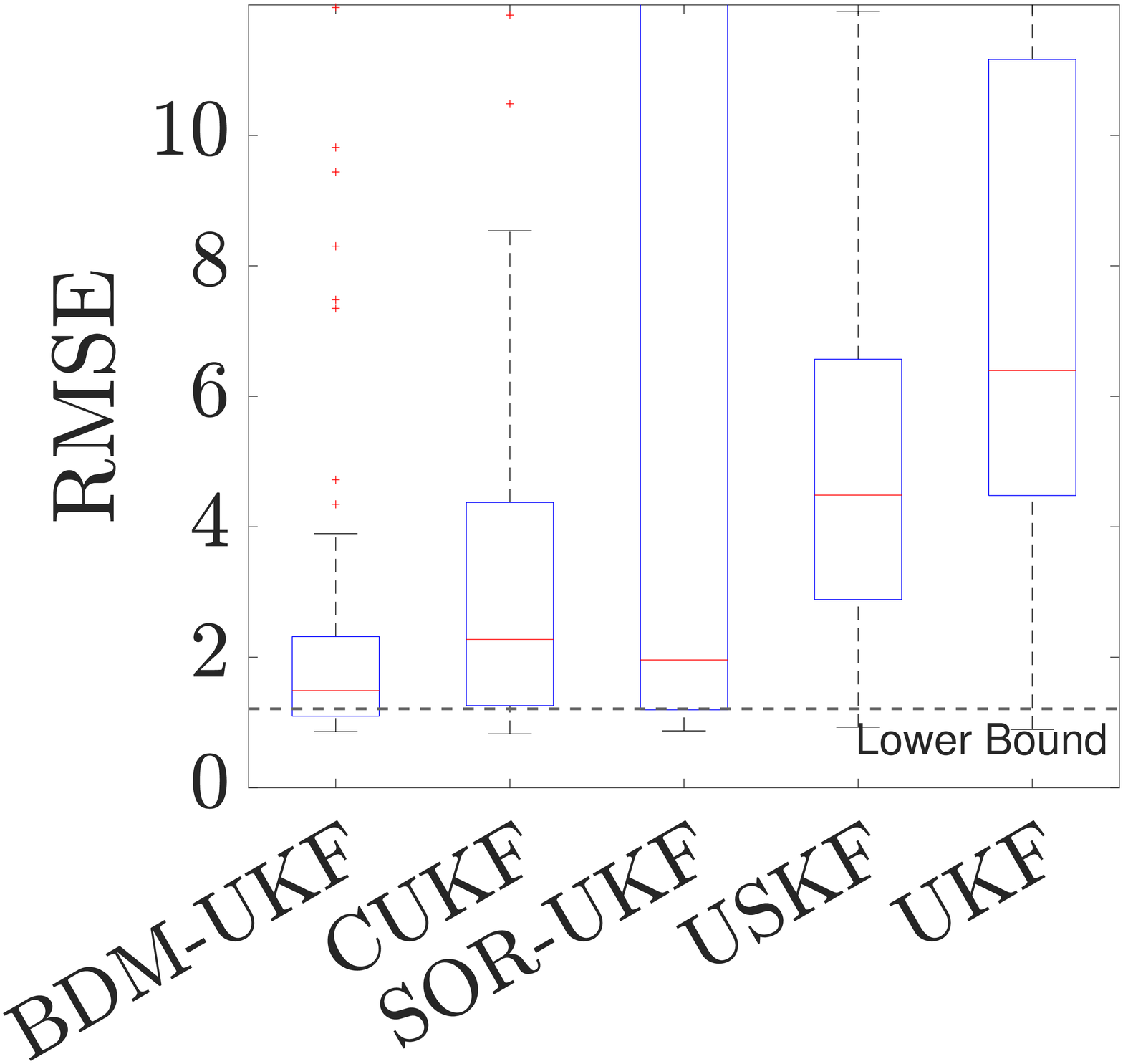}
		\caption{$\lambda = 0.6$}
		\label{Box23}
	\end{subfigure}
	\begin{subfigure}[t!]{0.22\textwidth}
		\centering
		\includegraphics[width=\linewidth,trim=0 0 20cm 0,clip=true]{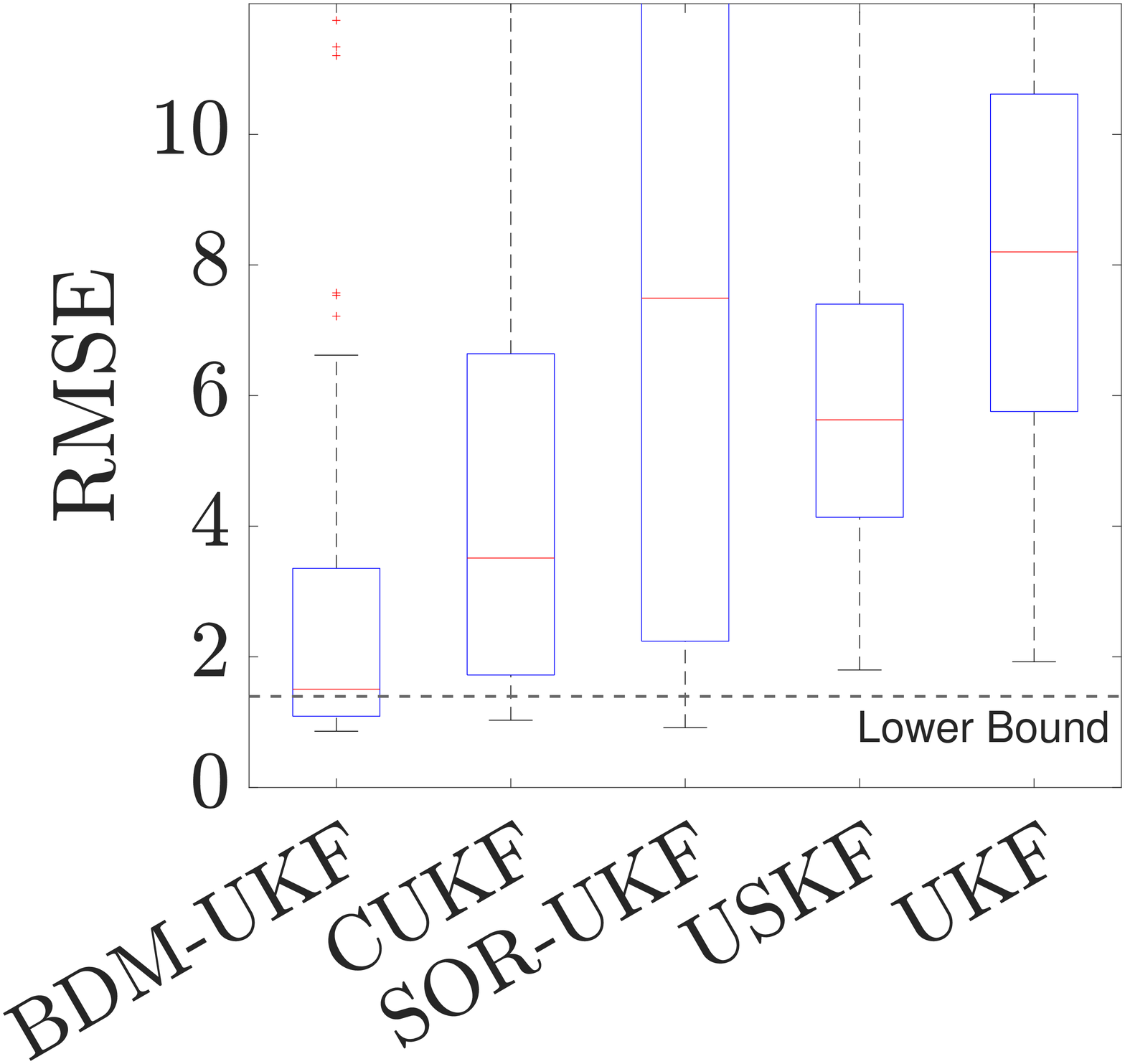}
		\caption{$\lambda = 0.8$}
		\label{Box24}
	\end{subfigure}
	\caption{Box plots of state RMSE for Case 2 with increasing $\lambda$}\label{Box2}
\end{figure*}

\begin{figure}[ht!]
	\centering
	\vspace{.07cm}
	\includegraphics[width=\linewidth]{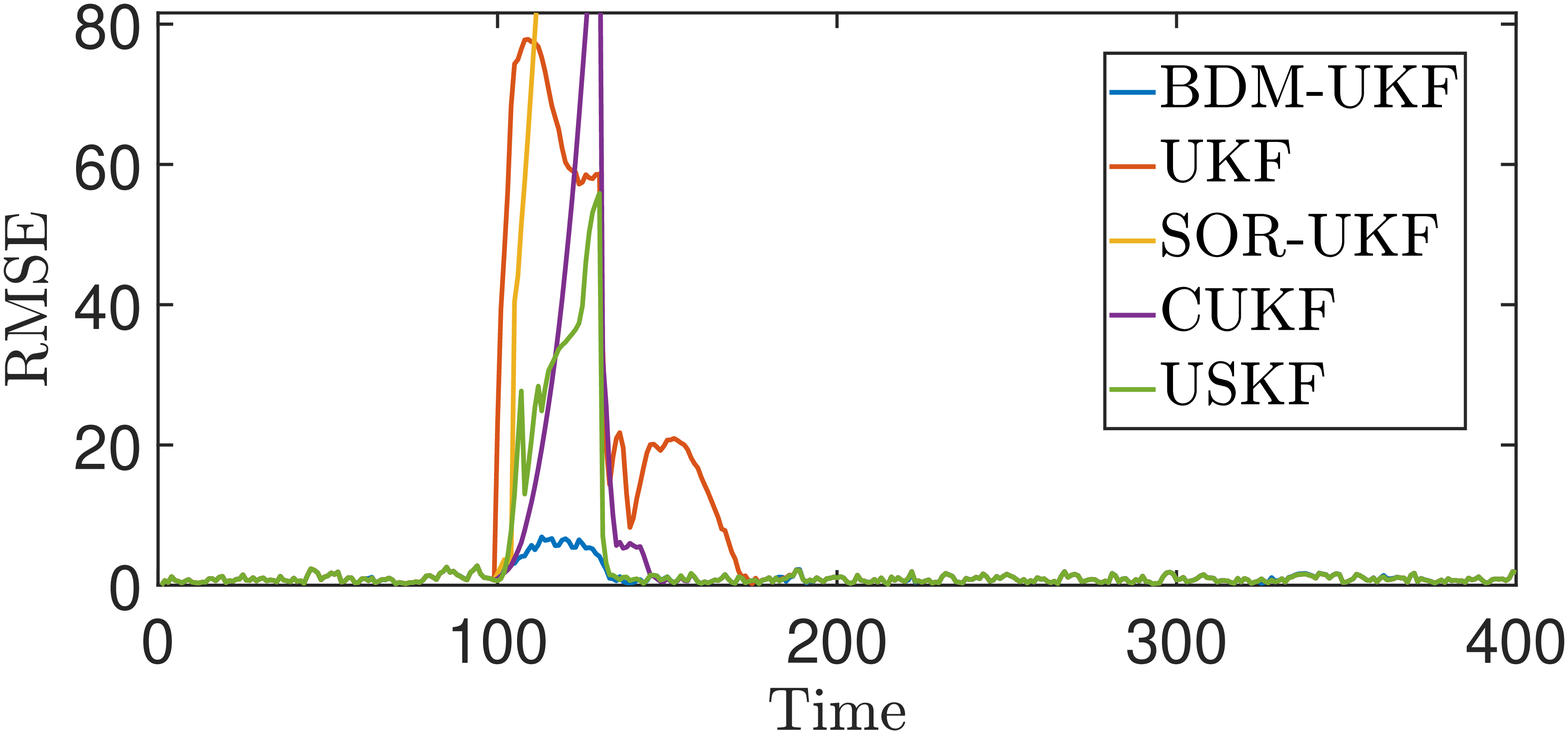}
	\caption{Tracking performance of algorithms over time for an example MC run for Case 2}
	\label{fig:Case2_run}
\end{figure}

\subsection{Case 1: Consistent bias presence}
\begin{table}[b!]
	\centering
	\begin{tabular}{|c c c c c c|} 
		\hline
		$\boldsymbol{\lambda}$   &\textbf{UKF}&\textbf{USKF}& \textbf{BDM-UKF} & \textbf{SOR-UKF} & \textbf{CUKF}  \\ 
		\hline\hline
		0.2  &0.0469&0.0781& 0.0866 & 0.0821& 0.5762  \\ 
		0.4  &0.0470&0.0849& 0.1000 & 0.1199& 1.2499 \\
		0.6 &0.0471&0.0883& 0.1121 & 0.1416& 1.8012  \\
		0.8 &0.0470&0.0903& 0.1492 & 0.1637& 2.2305  \\ 
		\hline
	\end{tabular}
	\caption{Average time for 100 MC runs with constant bias at different values of $\lambda$ .}
	\label{table:1}
\end{table}

First, we consider the case of biases consistently appearing in the measurements representative of real-world scenarios where the observations are systematically biased. We consider that the bias corrupts each observation according to \eqref{eqn_res2c}-\eqref{eqn_res2d} and bias occurs in each dimension from the start of the simulation with probability $\lambda$ and sustains for the complete run time. 

Fig.~\ref{Box1} shows the distribution of the root mean squared error (RMSE) calculated over 100 MC runs for the algorithms under consideration at different values of $\lambda$. {The lower bounds based on the diagonal entries of $\text{PCRB}_k$ depict the benchmark performance for different methods theoretically achievable considering different values of $\lambda$. For evaluation of $\text{PCRB}_k$ we assume $N_{mc1}=N_{mc2}=N_{mc3}=N_{mc4}=100$.} We can observe an intuitive trend in the relative performance of different methods. The standard UKF exhibits the worst estimation quality since it has no means for bias detection and compensation. For the data rejection-based method i.e. the SOR-UKF, we see that for lower probabilities of existence of bias discounting the observations performs satisfactorily. However, for higher values of $\lambda$ the rejection scheme does not work well since consistently rejecting large number of the measurements leads to loss of essential information. The CUKF and the USKF generally perform better than the SOR-UKF. \textcolor{black}{As compared to the other robust filters} the USKF is found to have more error at lower values of $\lambda$ since it does not treat each dimension selectively.  It rather uses the entire vector of measurements for compensation even if one of the dimensions is corrupted unlike the CUKF which offers a selective treatment. Importantly, it can be observed that BDM-UKF results in the least error among all the methods. 

Fig.~\ref{fig:Case1_run} shows the state RMSE of the algorithms over time for one MC run depictive of the general trend in Fig.\ref{Box1} for large values of $\lambda$ for this case. The BDM-UKF deals with biased observations more effectively followed by the partially compensating methods namely the CUKF and the USKF. The SOR-UKF, on the other hand, loses track of the ground truth due to permanent rejection of essential information. Similarly, the standard UKF with its inability to deal with \textcolor{black}{biases} also exhibits large errors.

Lastly, we evaluate the computational overhead of each algorithm for Case 1. The mean processing time, considering 100 MC runs, for each method has been summarized in Table \ref{table:1}. We find the standard UKF to be the most economical and does not exhibit any significant change with different values of $\lambda$. The USKF takes more time since it involves updating the state and bias correlation terms. We observe a rise in the processing time of the USKF with increasing $\lambda$ since the USKF is invoked more frequently than the standard UKF we utilize when no bias is detected. The BDM-UKF and the SOR-UKF, both having an inbuilt detection mechanism, have a similar processing burden. Lastly, we find the CUKF to be the most computationally expensive algorithm. This can be owed to the presence of a convex quadratically constrained quadratic program (QCQP) which we solve using the MATLAB optimization toolbox.

\subsection{Case 2: Momentary bias presence}
We also consider the case where bias randomly appears for a short duration, characterizing practical scenarios e.g. where ambient effects disturb the data signals briefly. We again use \eqref{eqn_res2c}-\eqref{eqn_res2d} for each observation contamination and suppose bias occurs in each dimension with probability $\lambda$ at instant $t=100$ and sustains till $t=130$ before disappearing.

Fig.~\eqref{Box2} depicts the spread of the RMSEs for 100 MC runs for each algorithm with varying $\lambda$ {along with lower bounds based on the PCRB using 100 MC samples for all the calculations. The plotting scale in this case allows us to clearly see a rise in the lower bound with increasing values of $\lambda$ which has also been observed for the previous case. This can be attributed to more chances of occurrence of \textcolor{black}{biased observations} with increasing $\lambda$ leading to increased estimation uncertainty}. In terms of the relative performance of different methods we see a similar pattern as for the previous case, owing to the same rationale regarding the functionality of the methods. The UKF generally has the largest RMSE, followed by the USKF, the SOR-UKF and the CUKF. The momentary appearance of bias does not degrade the performance of SOR-UKF as compared to the last case, except for $\lambda=0.8$ where it mostly diverges. SUKF is found to be relatively less effective at lower values of $\lambda$ due to its non-selective nature. Lastly, the BDM-UKF results in the lowest RMSE. Fig.~\ref{fig:Case2_run} shows the state RMSE of the algorithms over time for one MC run representative of the general trend in Fig.~\ref{Box2} for large values of $\lambda$ for this case. We can observe similar results as in the preceding case.

\begin{table}[h!]
	\centering
	\begin{tabular}{|c c c c c c|} 
		\hline
		$\boldsymbol{\lambda}$ &\textbf{UKF}&\textbf{USKF}& \textbf{BDM-UKF} & \textbf{SOR-UKF} & \textbf{CUKF}  \\ 
		\hline\hline
		0.2 &0.0458&0.0515& 0.0682 & 0.0669& 0.09114  \\ 
		0.4  &0.0489&0.0521& 0.0687 & 0.0672& 0.1110 \\
		0.6 &0.0482&0.0520& 0.0687 & 0.0828& 0.1470  \\
		0.8 &0.0478&0.0520& 0.0690 & 0.0902& 0.2059  \\ 
		\hline
	\end{tabular}
	\caption{Average time for 100 MC runs \textcolor{black}{with momentary bias} at different values of $\lambda$ .}
	\label{table:2}
\end{table}

Lastly, the mean processing overhead of each algorithm for this case is presented in Table \ref{table:2}. We can see that the UKF takes approximately similar times as for Case 1. The overhead of remaining methods is reduced since the bias duration has now decreased. The order in which different algorithms appear in terms of the relative computational expense remains the same as observed in the previous case following from the same reasoning.

\section{Conclusion}\label{Conc}
The performances of standard filtering approaches degrade when the measurements are disturbed by noise with unknown statistics. Focusing on the presence of measurement \textcolor{black}{biases}, we devise the BDM filter with inherent error detection and mitigation functionality. Performance evaluation reveals the efficacy of the BDM in dealing with both persistently and temporarily present biases. We find the BDM filter more accurate compared to rejection-based KF methods i.e. SOR-UKF. Moreover, owing to better utilization of the measurements, the BDM filter has lower estimation errors as compared to the methods with similar KF based approaches, the USKF and the CUKF, aiming to exploit information from the corrupted dimensions. The BDM filter is easier to employ as it avoids the use of external detectors and any optimization solver. The gains come at the expense of increased computational overhead which is comparatively higher compared to the UKF and USKF. However, it is comparable to SOR-UKF and lower than the CUKF which requires an additional optimization solver.

%

\appendix
\subsection{Predicting parameters of $\mathcal{N}(\mathbf{\Theta}_{k}|\mathbf{\hat{\Theta}}^-_{k},\mathbf{{\Sigma}}^-_{k})$}
\begin{flalign}
	\hat{\mathbf{\Theta}}_{k}^{-} &=  \langle {\mathbf{\Theta}}_{k} \rangle_{p(\bm{{\Theta}}_{k}|\mathbf{y}_{1:{k\text{-}1}})}=\boldsymbol{\Omega}_{k\text{-}1}\hat{\mathbf{\Theta}}_{k\text{-}1}^{+}\\
	\mathbf{\Sigma}_{k}^{-} &=\langle \mathbf{v_1}_k \mathbf{v_1}^{\top}_k \rangle_{\begin{subarray} {l}  p(\bm{\mathcal{I}}_{k\text{-}1}|\mathbf{y}_{1:{k\text{-}1}}).p(\widetilde{\bm{\Theta}}_k).  p(\bm{{\Theta}}_{k\text{-}1}|\mathbf{y}_{1:{k\text{-}1}}). 
	p({\Delta_k})	\end{subarray}} &
\end{flalign}
where
\begin{flalign}
	\mathbf{v_1}_k&= (\mathbf{I}-\boldsymbol{\mathcal{I}}_{k\text{-}1})\widetilde{\mathbf{\Theta}}_{k} + \boldsymbol{\mathcal{I}}_{k\text{-}1}(\mathbf{\Theta}_{k\text{-}1}+{\Delta_k})-\boldsymbol{\Omega}_{k\text{-}1}\hat{\mathbf{\Theta}}_{k\text{-}1}^{+} & \nonumber \\
	&\hspace{.5cm}-\boldsymbol{\mathcal{I}}_{k\text{-}1}\hat{\mathbf{\Theta}}_{k\text{-}1}^{+}+\boldsymbol{\mathcal{I}}_{k\text{-}1}\hat{\mathbf{\Theta}}_{k\text{-}1}^{+}&
\end{flalign}
Since $\langle \widetilde{\mathbf{\Theta}}_{k}\rangle_{p(\widetilde{\bm{\Theta}}_k)}=\langle {\Delta_k} \rangle_{p({\Delta_k})}=\langle \mathbf{\Theta}_{k\text{-}1}\text{-}\hat{\mathbf{\Theta}}_{k\text{-}1}^{+} \rangle_{p(\bm{{\Theta}}_{k\text{-}1}|\mathbf{y}_{1:{k\text{-}1}})}=\mathbf{0}$, other terms in the expression of $\mathbf{\Sigma}_{k}^{-}$ disappear and we can write
	

\begin{flalign}
	\mathbf{\Sigma}_{k}^{-}&={\langle} (\mathbf{I}-\boldsymbol{\mathcal{I}}_{k\text{-}1})\widetilde{\mathbf{\Theta}}_{k}\widetilde{\mathbf{\Theta}}_{k}^{\top}(\mathbf{I}-\boldsymbol{\mathcal{I}}_{k\text{-}1})^{\top} + \boldsymbol{\mathcal{I}}_{k\text{-}1}{\Delta_k} {\Delta^{\top}_k}\boldsymbol{\mathcal{I}}_{k\text{-}1}^{\top} & \nonumber \\
	&+ \boldsymbol{\mathcal{I}}_{k\text{-}1}(\mathbf{\Theta}_{k\text{-}1} - \hat {\mathbf{\Theta}}_{k\text{-}1}^{+})(\mathbf{\Theta}_{k\text{-}1} -{{}\hat{\mathbf{\Theta}}_{k\text{-}1}^{+}})^{\top}\boldsymbol{\mathcal{I}}_{k\text{-}1}^{\top} & \nonumber \\
	& + (\boldsymbol{\mathcal{I}}_{k\text{-}1} - \boldsymbol{\Omega }_{k\text{-}1})\hat{\mathbf{\Theta}}_{k\text{-}1}^{+} {{}\hat{\mathbf{\Theta}}_{k\text{-}1}^{+}}^{\top}(\boldsymbol{\mathcal{I}}_{k\text{-}1} - \boldsymbol{\Omega}_{k\text{-}1})^{\top} {\rangle_{\begin{subarray} {l}  p(\bm{\mathcal{I}}_{k\text{-}1}|\mathbf{y}_{1:{k\text{-}1}}).\\  p(\bm{{\Theta}}_{k\text{-}1}|\mathbf{y}_{1:{k\text{-}1}}).\\p(\widetilde{\bm{\Theta}}_k). 
	p({\Delta_k})	\end{subarray}} }& 
\end{flalign}
We can further write
\begin{flalign}
	\mathbf{\Sigma}_{k}^{-}&=\langle \mathbf{v_2}_k \mathbf{v_2}^{\top}_k+\mathbf{v_3}_k \mathbf{v_3}^{\top}_k+\mathbf{v_4}_k \mathbf{v_4}^{\top}_k+\mathbf{v_5}_k \mathbf{v_5}^{\top}_k \rangle_{\begin{subarray} {l}  p(\bm{\mathcal{I}}_{k\text{-}1}|\mathbf{y}_{1:{k\text{-}1}}).\\  p(\bm{{\Theta}}_{k\text{-}1}|\mathbf{y}_{1:{k\text{-}1}}).\\p(\widetilde{\bm{\Theta}}_k). 
	p({\Delta_k})	\end{subarray}}  &
\end{flalign}
with $\mathbf{v_2}_k=\begin{pmatrix}
	(1\text{-}{\mathcal{I}}^{1}_{k\text{-}1})\widetilde{\Theta} _{k}^{1} \\
	\vdots\\
	(1\text{-}{\mathcal{I}}^{m}_{k\text{-}1})\widetilde{\Theta} _{k}^{m}
\end{pmatrix},\mathbf{v_3}_k=\begin{pmatrix}
{\mathcal{I}}^{1}_{k\text{-}1} {\Delta^1_k} \\
\vdots\\
{\mathcal{I}}^{m}_{k\text{-}1}{\Delta^m_k} 
\end{pmatrix}$ \\

$\mathbf{v_4}_k=\begin{pmatrix}
	{\mathcal{I}}^{1}_{k\text{-}1}(\Theta_{k\text{-}1}^{1} \text{-} {{}\hat{\Theta}_{k\text{-}1}^{+^1}}) \\
	\vdots\\
	\mathcal{I}^{m}_{k\text{-}1}({\Theta^{m}_{k\text{-}1}} \text{-} {{}\hat{\Theta}_{k\text{-}1}^{+^m}})
\end{pmatrix},\mathbf{v_5}_k=\begin{pmatrix}
(\mathcal{I}_{k\text{-}1}^{1} \text{-} \Omega_{k\text{-}1}^{1}) {{}\hat{\Theta}_{k\text{-}1}^{+^1}} \\
\vdots\\
(\mathcal{I}_{k\text{-}1}^{m} \text{-} \Omega _{k\text{-}1}^{m}) {{}\hat{\Theta}_{k\text{-}1}^{+^{m}}}
\end{pmatrix}$
\newline
\newline
Resultingly $\mathbf{\Sigma}_{k}^{-}$ can be further expressed as
\begin{flalign}
	&\mathbf{\Sigma}_{k}^{-}\text{=} \langle 
	\begin{pmatrix}
		{(1\text{-}\mathcal{I}^{1}_{k\text{-}1})^{2}{{} \widetilde{\Theta} _{k}^{1}}^{2}} & \cdots & \Pi_{i} (1\text{-}\mathcal{I}^{i}_{k\text{-}1})\widetilde{\Theta} _{k}^{i}\\
		\vdots & \ddots & \vdots \\
		. & \cdots & (1\text{-}\mathcal{I}^{m}_{k\text{-}1})^{2}{{}\widetilde{\Theta} _{k}^{m}}^{2}
	\end{pmatrix}
	 \rangle_{{\begin{subarray}{l}		p(\bm{\mathcal{I}}_{k\text{-}1}|\mathbf{y}_{1:{k\text{-}1}}).\\ p(\bm{\widetilde{\Theta}}_{k})  
	 		\end{subarray}
	 }} & \nonumber \\
	& \text{+} \langle 
	\begin{pmatrix}
		{\mathcal{I}_{k\text{-}1}^{1}}^{2}{\Delta^1_k}^2 & \cdots & \Pi_{i} {\mathcal{I}_{k\text{-}1}^{i}} {\Delta^i_k} \\
		\vdots & \ddots & \vdots \\
		. & \cdots & {\mathcal{I}_{k\text{-}1}^{m}}^2 {\Delta^m_k}^2
	\end{pmatrix}
	\rangle_{{\begin{subarray}{l}		p(\bm{\mathcal{I}}_{k\text{-}1}|\mathbf{y}_{1:{k\text{-}1}}). p(\Delta_k)  
			\end{subarray}
	}} & \nonumber \\
	&\text{+} \langle 
	\begin{pmatrix}
		{\mathcal{I}^{1^2}_{k\text{-}1}}(\Theta_{k\text{-}1}^{1} \text{-} {{}\hat{\Theta}_{k\text{-}1}^{\text{+}^{1}}})^{2} & \cdots & \Pi_{i} {\mathcal{I}^{i}_{k\text{-}1}}(\Theta_{k\text{-}1}^{i} \text{-} {{}\hat{\Theta}_{k\text{-}1}^{\text{+}^{i}}}) \nonumber \\
		\vdots & \ddots & \vdots \\
		. & \cdots & {\mathcal{I}^{m^2}_{k\text{-}1}}(\Theta_{k\text{-}1}^{m} \text{-} {{}\hat{\Theta}_{k\text{-}1}^{\text{+}^{m}}})^{2}
	\end{pmatrix}
	\rangle_{{\begin{subarray}{l}		p(\bm{\mathcal{I}}_{k\text{-}1}|\mathbf{y}_{1:{k\text{-}1}}).\\  p(\bm{{\Theta}}_{k\text{-}1}|\mathbf{y}_{1:{k\text{-}1}})  
\end{subarray}
		}} &
\\
	&\text{+}\langle 
	\begin{pmatrix}
		{(\mathcal{I}_{k\text{-}1}^{1} \text{-} \Omega _{k\text{-}1}^{1})}^{2}  {{{}\hat {\Theta}_{k\text{-}1}^{\text{+}^{1}}}}^{2} & {\cdots} & \Pi_{i} {(\mathcal{I}_{k\text{-}1}^{i} \text{-} \Omega _{k\text{-}1}^{i})}  {{{}\hat {\Theta}_{k\text{-}1}^{\text{+}^{i}}}} \\
		\vdots & {\ddots} & \vdots \\
		. & {\cdots} & {(\mathcal{I}_{k\text{-}1}^{m} \text{-} \Omega _{k\text{-}1}^{m})}^{2}  {{{}\hat {\Theta}_{k\text{-}1}^{\text{+}^{m}}}}^{2}
	\end{pmatrix}
	\rangle_{{\begin{subarray}{l}		p(\bm{\mathcal{I}}_{k\text{-}1}|\mathbf{y}_{1:{k\text{-}1}})\end{subarray}
	}}&
\end{flalign}
where ${i\in\{1,m\}}$ 
\begin{flalign}
	&{=}\mathrm{diag}	\begin{pmatrix}\langle (1\text{-}\mathcal{I}^{1}_{k\text{-}1})^{2}\rangle_{p(\bm{\mathcal{I}}_{k\text{-}1}|\mathbf{y}_{1:{k\text{-}1}})}\\
		\vdots\\\langle(1\text{-}\mathcal{I}^{m}_{k\text{-}1})^{2}\rangle_{p(\bm{\mathcal{I}}_{k\text{-}1}|\mathbf{y}_{1:{k\text{-}1}})}
		\end{pmatrix} \mathrm{diag}	\begin{pmatrix}\langle { {}\widetilde{\Theta} _{k}^{1}}^{2} \rangle_{p(\bm{\widetilde{\Theta}}_{k})} \\
		\vdots\\\langle{ {}\widetilde{\Theta} _{k}^{m}}^{2}\rangle_{p(\bm{\widetilde{\Theta}}_{k})} \end{pmatrix} & \\
	&{+}\mathrm{diag}	\begin{pmatrix}\langle {\mathcal{I}^{1}_{k\text{-}1}}^{2}\rangle_{p(\bm{\mathcal{I}}_{k\text{-}1}|\mathbf{y}_{1:{k\text{-}1}})}\\
			\vdots\\\langle{\mathcal{I}^{m}_{k\text{-}1}}^{2}\rangle_{p(\bm{\mathcal{I}}_{k\text{-}1}|\mathbf{y}_{1:{k\text{-}1}})}
		\end{pmatrix} \mathrm{diag}	\begin{pmatrix}\langle {\Delta^1_k}^2  \rangle_{p(\Delta_{k})} \\
			\vdots\\ \langle {\Delta^m_k}^2 \rangle_{p(\Delta_{k})}
		\end{pmatrix}+ \mathbf{A}_{k\text{-}1}&	\\
    &{+}\mathrm{diag}	\begin{pmatrix}\langle {(\mathcal{I}_{k\text{-}1}^{1} \text{-} \Omega _{k\text{-}1}^{1})^2}\rangle_{p(\bm{\mathcal{I}}_{k\text{-}1}|\mathbf{y}_{1:{k\text{-}1}})}\\
		\vdots\\\langle {(\mathcal{I}_{k\text{-}1}^{m} \text{-} \Omega _{k\text{-}1}^{m})^2}\rangle_{p(\bm{\mathcal{I}}_{k\text{-}1}|\mathbf{y}_{1:{k\text{-}1}})}
	\end{pmatrix} \mathrm{diag}	\begin{pmatrix} {{{}\hat {\Theta}_{k\text{-}1}^{+^1}}}^{2} \\
		\vdots\\ {{{}\hat {\Theta}_{k\text{-}1}^{+^m}}}^{2} 
	\end{pmatrix} &
\end{flalign}
where
\begin{flalign}
\langle (1\text{-}\mathcal{I}^{i}_{k\text{-}1})^{2}\rangle_{p(\bm{\mathcal{I}}_{k\text{-}1}|\mathbf{y}_{1:{k\text{-}1}})}&=(1-\Omega^{+^i}_{k\text{-}1})&\\
\langle { {}\widetilde{\Theta} _{k}^{i}}^{2} \rangle_{p(\bm{\widetilde{\Theta}}_{k})} &=\widetilde{{\Sigma}}^i_{k}&\\
\langle {\mathcal{I}^{i}_{k\text{-}1}}^{2}\rangle_{p(\bm{\mathcal{I}}_{k\text{-}1}|\mathbf{y}_{1:{k\text{-}1}})}&=\Omega^{+^i}_{k\text{-}1}&\\
\langle {\Delta^i_k}^2 \rangle_{p(\Delta_{k})}&=\breve{{\Sigma}}^i_{k}&
\end{flalign}
\begin{flalign}
	{A}^{ij}_{k\text{-}1}&=\langle 
	{\mathcal{I}^{i}_{k\text{-}1}}{\mathcal{I}^{j}_{k\text{-}1}}(\Theta_{k\text{-}1}^{i} {-} {{}\hat{\Theta}_{k\text{-}1}^{\text{+}^{i}}})(\Theta_{k\text{-}1}^{j} {-} {{}\hat{\Theta}_{k\text{-}1}^{\text{+}^{j}}})\rangle_{{\begin{subarray}{l}		p(\bm{\mathcal{I}}_{k\text{-}1}|\mathbf{y}_{1:{k\text{-}1}}).\\  p(\bm{{\Theta}}_{k\text{-}1}|\mathbf{y}_{1:{k\text{-}1}})  
			\end{subarray}
	}}&
\end{flalign}
for $i=j$
\begin{flalign}
	{A}^{ij}_{k\text{-}1}&=\langle 
	{\mathcal{I}^{i}_{k\text{-}1}}^2 \rangle_{{\begin{subarray}{l}		p(\bm{\mathcal{I}}_{k\text{-}1}|\mathbf{y}_{1:{k\text{-}1}})
			\end{subarray}
	}} \langle (\Theta_{k\text{-}1}^{i} {-} {{}\hat{\Theta}_{k\text{-}1}^{\text{+}^{i}}})^2\rangle_{{\begin{subarray}{l}		p(\bm{{\Theta}}_{k\text{-}1}|\mathbf{y}_{1:{k\text{-}1}})  
	\end{subarray}
}} &\nonumber \\
&=\Omega^{+^i}_{k\text{-}1} \Sigma^{+^{ii}}_{k\text{-}1}&
\end{flalign}
for $i \neq j$
\begin{flalign}
	{A}^{ij}_{k\text{-}1}&=\langle 
	{\mathcal{I}^{i}_{k\text{-}1}}\rangle_{{\begin{subarray}{l}		p(\bm{\mathcal{I}}_{k\text{-}1}|\mathbf{y}_{1:{k\text{-}1}})
			\end{subarray}
	}} \langle {\mathcal{I}^{j}_{k\text{-}1}}\rangle_{{\begin{subarray}{l}	p(\bm{\mathcal{I}}_{k\text{-}1}|\mathbf{y}_{1:{k\text{-}1}})
	\end{subarray}
}} &\nonumber \\ 
&\ \ \times \langle (\Theta_{k\text{-}1}^{i} {-} {{}\hat{\Theta}_{k\text{-}1}^{\text{+}^{i}}})(\Theta_{k\text{-}1}^{j} {-} {{}\hat{\Theta}_{k\text{-}1}^{\text{+}^{j}}})\rangle_{{\begin{subarray}{l}		p(\bm{{\Theta}}_{k\text{-}1}|\mathbf{y}_{1:{k\text{-}1}})  
			\end{subarray}
	}}	&\nonumber \\
&=\Omega^{+^i}_{k\text{-}1} \Omega^{+^j}_{k\text{-}1} \Sigma^{+^{ij}}_{k\text{-}1}&
\end{flalign}
$\therefore \mathbf{A}_{k\text{-}1}$ can be written as
\begin{flalign}
\mathbf{A}_{k\text{-}1}&=\mathbf{{\Sigma}}^+_{k\text{-}1}\odot(\mathrm{diag}(\mathbf{\Omega}_{k\text{-}1}){\mathrm{diag}(\mathbf{\Omega}_{k\text{-}1})}^{\top}+\mathbf{\Omega}_{k\text{-}1}(\mathbf{I}-\mathbf{\Omega}_{k\text{-}1}))&	\\
\mathbf{\Sigma}^{-}_{k}&=(\mathbf{I}-\mathbf{\Omega}_{k\text{-}1})\widetilde{\mathbf{\Sigma}}_{k}+\mathbf{\Omega}_{k\text{-}1}\breve{\mathbf{\Sigma}}_{k}&\nonumber\\
&+\mathbf{{\Sigma}}^+_{k\text{-}1}\odot(\mathrm{diag}(\mathbf{\Omega}_{k\text{-}1}){\mathrm{diag}(\mathbf{\Omega}_{k\text{-}1})}^{\top}+\mathbf{\Omega}_{k\text{-}1}(\mathbf{I}-\mathbf{\Omega}_{k\text{-}1}))&\nonumber \\
&+\mathbf{\Omega}_{k\text{-}1}(\mathbf{I}-\mathbf{\Omega}_{k\text{-}1})(\mathrm{diag}(\mathbf{\hat{\Theta}}^{+}_{k\text{-}1}))^2 &	
\end{flalign}

\subsection{Derivation of $q(\mathbf{x}_{k})$}
Using \eqref{eqn_vb_3} and \eqref{eqn_vb_8} we can write $q(\mathbf{x}_k)$ as
\begin{flalign}
	&q(\mathbf{x}_k)\propto \exp\big( \big\langle\mathrm{ln}(p(\mathbf{y}_k|\mathbf{x}_{k},\bm{\mathcal{I}}_k,{\mathbf{\Theta}}_k) p(\bm{\mathcal{I}}_k)p(\mathbf{x}_k|\mathbf{y}_{1:k\text{-}1}) &\nonumber \\ &\hspace{1.2cm}p({\mathbf{\Theta}}_k|\mathbf{y}_{1:k\text{-}1})) \rangle_{ q({{\bm{\mathcal{I}}}_k}) {q(\mathbf{\Theta}}_k)}\big)& \\
	&\propto \exp\big( \big\langle\mathrm{ln}(p(\mathbf{y}_k|\mathbf{x}_{k},\bm{\mathcal{I}}_k,{\mathbf{\Theta}}_k)\rangle_{ q({{\bm{\mathcal{I}}}_k}) {q(\mathbf{\Theta}}_k)}\big) p(\mathbf{x}_k|\mathbf{y}_{1:k\text{-}1}) &\\
	&\propto \exp \big( \langle \mathrm{ln}(\mathcal{N}(\mathbf{y}_{k}|\mathbf{h}(\mathbf{x}_{k}) + \boldsymbol{\mathcal{I}}_{k}\mathbf{\Theta}_{k},\mathbf{R}_{k})\rangle_{ q({{\bm{\mathcal{I}}}_k}) {q(\mathbf{\Theta}}_k)}\big)\nonumber\\&\hspace{.5cm}\mathcal{N}(\mathbf{x}_{k}|\hat {\mathbf{x}}_{k}^{-},\mathbf{P}_{k}^{-}) &
	\\
	&\propto \exp ( \langle (-\frac{1}{2}\mathbf{v_6}^{\top}_k 
	 \mathbf{R}_{k}^{-1} \mathbf{v_6}_k ) -\frac{1}{2} \mathrm{ln}(2 \pi) ^{m} | \mathbf{R}_{k}|)\rangle_{ q({{\bm{\mathcal{I}}}_k}) {q(\mathbf{\Theta}}_k)}\big)  & \nonumber 
	\\ 
	& \hspace{.5cm}\mathcal{N}(\mathbf{x}_{k}|\hat {\mathbf{x}}_{k}^{-},\mathbf{P}_{k}^{-})&\\
	&\propto \exp \big(-\frac{1}{2} \langle \mathrm{tr} (\mathbf{v_6}_k\mathbf{v_6}^{\top}_k 
	\mathbf{R}_{k}^{-1} )\rangle_{ q({{\bm{\mathcal{I}}}_k}) {q(\mathbf{\Theta}}_k)}\big) \mathcal{N}(\mathbf{x}_{k}|\hat {\mathbf{x}}_{k}^{-},\mathbf{P}_{k}^{-})&
\end{flalign}
where $\mathbf{v_6}_k=\mathbf{y}_{k}-\mathbf{h}(\mathbf{x}_{k})  - \boldsymbol{\mathcal{I}}_{k}\mathbf{\Theta}_{k}$. Furthermore, we can write
\begin{flalign}
		& q(\mathbf{x}_k)\propto
		\exp\big({-}\frac{1}{2} \mathrm{tr}\big((\mathbf{v_7}_k\mathbf{v_7}^{\top}_k- \mathbf{v_7}_k  {\mathbf{v_8}_k}^{\top}  - \mathbf{v_8}_k \mathbf{v_7}^{\top}_k ) \mathbf{R}_{k}^{-1}\big)\big) &\nonumber\\ &\hspace{1.2cm} \mathcal{N}(\mathbf{x}_{k}|\hat {\mathbf{x}}_{k}^{-},\mathbf{P}_{k}^{-})&
\end{flalign}
where $\mathbf{v_7}_k=\mathbf{y}_{k}-\mathbf{h}(\mathbf{x}_{k})$ and $\mathbf{v_8}_k=\mathbf{\Omega}_{k} {{}\hat{\mathbf{\Theta}}_{k}^{+}}$. Adding a constant term $\mathbf{v_8}_k\mathbf{v_8}_k^{\top}$ to complete the square in the exponential expression yields
\begin{flalign}
		&q(\mathbf{x}_k)\propto \exp \big(-\frac{1}{2}   (\mathbf{v_9}^{\top}_k 
	\mathbf{R}_{k}^{-1} \mathbf{v_9}_k)\big) \mathcal{N}(\mathbf{x}_{k}|\hat {\mathbf{x}}_{k}^{-},\mathbf{P}_{k}^{-})&
\end{flalign}
where $\mathbf{v_9}_k=\mathbf{y}_{k}-\mathbf{h}(\mathbf{x}_{k})  - \mathbf{\Omega}_{k} {{}\hat{\mathbf{\Theta}}_{k}^{+}}$. Using general Gaussian filtering results \cite{sarkka2013bayesian}, we can further write $q(\mathbf{x}_k)$ as
\begin{flalign}
	&q(\mathbf{x}_k)\propto\mathcal{N}(\mathbf{x}_{k}|\mathbf{x}_{k}^{+},\mathbf{P}_{k}^{+})&
\end{flalign}
The parameters $\mathbf{x}_{k}^{+}$ and $\mathbf{P}_{k}^{+}$ can be updated using \eqref{eqn_vb_18}-\eqref{eqn_vb_23}. 

\subsection{\texorpdfstring{Derivation of $q(\boldsymbol{\mathcal{I}}_k)$}{}}
Using \eqref{eqn_vb_4} and \eqref{eqn_vb_8} we can write $q(\boldsymbol{\mathcal{I}}_{k})$ as
\begin{flalign}
	q(\boldsymbol{\mathcal{I}}_{k})\propto &\exp\big( \big\langle\mathrm{ln}(p(\mathbf{y}_k|\mathbf{x}_{k},\bm{\mathcal{I}}_k,{\mathbf{\Theta}}_k) p(\bm{\mathcal{I}}_k)p(\mathbf{x}_k|\mathbf{y}_{1:k\text{-}1}) &\nonumber \\ &p({\mathbf{\Theta}}_k|\mathbf{y}_{1:k\text{-}1})) \rangle_{ q(\mathbf{x}_k) {q(\mathbf{\Theta}}_k)}\big)& \\
	\propto &\exp\big( \big\langle\mathrm{ln}(p(\mathbf{y}_k|\mathbf{x}_{k},\bm{\mathcal{I}}_k,{\mathbf{\Theta}}_k) \rangle_{ q(\mathbf{x}_k) {q(\mathbf{\Theta}}_k)}\big) p(\bm{\mathcal{I}}_k)&  
\end{flalign}
As we consider $\textbf{R}_k$ to be diagonal we can write
\begin{flalign}
	q(\boldsymbol{\mathcal{I}}_{k})\propto 
	&\exp\big( \sum_i -\frac{1}{2 R_{k}^{i}} a_k \big) \begin{pmatrix}
		\prod_{i} (1-\theta_{k}^{i})\delta(\mathcal{I}_{k}^{i}) + \theta_{k}^{i}(\mathcal{I}_{k}^{i} - 1)\end{pmatrix}  &
\end{flalign}
where 
\begin{align}
	&a_k=\langle {({y}_{k}^{i} - (h^{i}(\mathbf{x}_{k}) +   \mathcal{I}^{i}_{k}\Theta^{i}_{k}))^{2}} \rangle_{q(\mathbf{x}_{k})q(\mathbf{\Theta}_{k})}\\
	& {=}\langle  {(h^{i}(\mathbf{x}_{k}) - {\nu}_k^{i} + {\mathcal{I}}_{k}({\Theta}_{k}^i - {{{}\hat {\Theta}_{k}^{\text{+}^{i}}}}) + {\nu}_k^{i} + {\mathcal{I}}_{k}{{{}\hat {\Theta}_{k}^{\text{+}^{i}}}} - y_{k}^{i})^{2}}  \rangle_{{\begin{subarray}{l}		q(\mathbf{x}_{k}).\\q(\mathbf{\Theta}_{k})  
			\end{subarray}
	}}  \\
	&{=}\bar{h}^2_k+{{\mathcal{I}}_{k}}^2 \bar{\Theta}^2_k+ ({\nu}_k^{i} + {\mathcal{I}}_{k}{{{}\hat {\Theta}_{k}^{\text{+}^{i}}}} - y_{k}^{i})^{2}  	
\end{align}
Consequently $q(\boldsymbol{\mathcal{I}}_{k})$ can be expressed as follows with its parameters updated using \eqref{eqn_vb_29}-\eqref{eqn_vb_28}.
\begin{flalign}
	&q(\boldsymbol{\mathcal{I}}_{k})
	=\prod_{i=1}^{m} (1-{\Omega^i_{k\text{-}1}}) \delta({{{\mathcal{I}}}^i_{k\text{-}1}})+{\Omega^i_{k\text{-}1}}\delta( {{{\mathcal{I}}}^i_{k\text{-}1}}-1)&
\end{flalign}

\subsection{Derivation of  $q(\mathbf{\Theta}_{k})$}
Using \eqref{eqn_vb_5} and \eqref{eqn_vb_8} we can write $q(\mathbf{\Theta}_k)$ as
\begin{flalign}
	&q(\mathbf{\Theta}_{k})\propto \exp\big( \big\langle\mathrm{ln}(p(\mathbf{y}_k|\mathbf{x}_{k},\bm{\mathcal{I}}_k,{\mathbf{\Theta}}_k) p(\bm{\mathcal{I}}_k)p(\mathbf{x}_k|\mathbf{y}_{1:k\text{-}1}) &\nonumber \\ &\hspace{1.2cm}p({\mathbf{\Theta}}_k|\mathbf{y}_{1:k\text{-}1})) \rangle_{ q(\mathbf{\mathbf{x}_{k}}) q({{\bm{\mathcal{I}}}_k}) }\big)& \\
	&\propto \exp\big( \big\langle\mathrm{ln}(p(\mathbf{y}_k|\mathbf{x}_{k},\bm{\mathcal{I}}_k,{\mathbf{\Theta}}_k)\rangle_{ q(\mathbf{\mathbf{x}_{k}}) q({{\bm{\mathcal{I}}}_k}) } \big) p(\mathbf{\Theta}_k|\mathbf{y}_{1:k\text{-}1}) &\\
	&\propto \exp \big( \langle \mathrm{ln}(\mathcal{N}(\mathbf{y}_{k}|\mathbf{h}(\mathbf{x}_{k}) + \boldsymbol{\mathcal{I}}_{k}\mathbf{\Theta}_{k},\mathbf{R}_{k})\rangle_{ q(\mathbf{\mathbf{x}_{k}}) q({{\bm{\mathcal{I}}}_k}) }\big)\nonumber\\&\hspace{.5cm}\mathcal{N}(\mathbf{\Theta}_{k}|\mathbf{\hat{\Theta}}^-_{k},\mathbf{{\Sigma}}^-_{k}) &
	\\
	&\propto \exp ( \langle (-\frac{1}{2}\mathbf{v_6}^{\top}_k 
	\mathbf{R}_{k}^{-1} \mathbf{v_6}_k ) -\frac{1}{2} \mathrm{ln}(2 \pi) ^{m} | \mathbf{R}_{k}|)\rangle_{ q(\mathbf{\mathbf{x}_{k}}) q({{\bm{\mathcal{I}}}_k}) }\big)  & \nonumber 
	\\ 
	& \hspace{.5cm}\mathcal{N}(\mathbf{\Theta}_{k}|\mathbf{\hat{\Theta}}^-_{k},\mathbf{{\Sigma}}^-_{k})&\\
	&\propto \exp \big(-\frac{1}{2} \langle \mathrm{tr} (\mathbf{v_6}_k\mathbf{v_6}^{\top}_k 
	\mathbf{R}_{k}^{-1} )\rangle_{ q({{\bm{\mathcal{I}}}_k}) {q(\mathbf{\Theta}}_k)}\big) \mathcal{N}(\mathbf{\Theta}_{k}|\mathbf{\hat{\Theta}}^-_{k},\mathbf{{\Sigma}}^-_{k}) &
\end{flalign}
where we write $\mathbf{v_6}_k$ in a useful form $\mathbf{v_6}_k=(\boldsymbol{\mathcal{I}}_{k}-\boldsymbol{\mathbf{\Omega}}_{k})\mathbf{\Theta}_{k}+\boldsymbol{\mathbf{\Omega}}_{k}\mathbf{\Theta}_{k}+\mathbf{h}(\mathbf{x}_{k})  -\mathbf{y}_{k}$ for the subsequent derivation of $q(\mathbf{\Theta}_{k})$ as
\begin{flalign}
	& q(\mathbf{\Theta}_k)\propto
	\exp\big({-}\frac{1}{2} \mathrm{tr}\big((\mathbf{B}_{k}+\mathbf{v_{10}}_k{\mathbf{v_{10}}_k}^{\top}+\mathbf{v_{10}}_k {\mathbf{v_{11}}_k}^{\top} &\nonumber \\  &\hspace{1.3cm} +\mathbf{v_{11}}_k {\mathbf{v_{10}}_k}^{\top})\mathbf{R}_{k}^{-1}\big)\big)\mathcal{N}(\mathbf{\Theta}_{k}|\mathbf{\hat{\Theta}}^-_{k},\mathbf{{\Sigma}}^-_{k})  &
\end{flalign}
where $\mathbf{B}_k=\mathrm{diag}(\mathbf{\Theta}_{k})\mathbf{\Omega}_k (\mathbf{I}-\mathbf{\Omega}_k)\mathrm{diag}(\mathbf{\Theta}_{k})$, $\mathbf{v_{10}}_k=\boldsymbol{\mathbf{\Omega}}_{k}\mathbf{\Theta}_{k}$ and $\mathbf{v_{11}}_k=\bm{\nu}_k-\mathbf{y}_{k}$ where $\bm{\nu}_k= \langle {\mathbf{h}}(\mathbf{x}_{k})  \rangle_{q(\mathbf{x}_k)}$. Adding a constant term $\mathbf{v_{11}}_k\mathbf{v_{11}}_k^{\top}$ to complete the square in the exponential expression and considering $\mathbf{R}_k$ to be diagonal yields
\begin{flalign}
	q(\mathbf{{\Theta}}_k)\propto
	&\overset{\mathcal{N}(\mathbf{\Theta}_{k}|\mathbf{0},(\boldsymbol{\Omega}_{k}(1-\boldsymbol{\Omega}_{k})\mathbf{R}_{k}^{-1})^{-1})}{\overbrace{\exp\big(-\frac{1}{2} \mathbf{\Theta}_{k}^{\top} (\boldsymbol{\Omega}_{k} (1 - \boldsymbol{\Omega}_{k})\mathbf{R}_{k}^{-1}\mathbf{\Theta}_{k}\big)}} \times & \nonumber \\ 
	&\underset{\mathcal{N}(\mathbf{\Theta}_{k}|\hat{\mathbf{\Theta}}_{k}^{*},\mathbf{\Sigma}_{k}^{*})}{\underbrace{\exp \big(-\frac{1}{2}   (\mathbf{v_{12}}^{\top}_k 
	\mathbf{R}_{k}^{-1} \mathbf{v_{12}}_k)\big) \mathcal{N}(\mathbf{\Theta}_{k}| \hat {\mathbf{\Theta}}_{k}^{-},\mathbf{\Sigma}_{k}^{-})}}& \\
	 \propto &\mathcal{N}\big( \mathbf{\Theta}_{k}| \hat{\mathbf{\Theta}}_{k}^{+}, \mathbf{\Sigma}_{k}^{+}\big)&
\end{flalign}
where $\mathbf{v_{12}}_k=\mathbf{y}_{k}-\bm{\nu}_k  - \mathbf{\Omega}_{k} {{\mathbf{\Theta}}_{k}}$. Using general Gaussian filtering results \cite{sarkka2013bayesian}, we can update $\hat {\mathbf{\Theta}}_{k}^{*}$ and $\mathbf{\Sigma}_{k}^{*} $ using \eqref{eqn_vb_30}-\eqref{eqn_vb_34}. The following appears as a result of the product of two multivariate Gaussian distributions \cite{petersen2008matrix}
\begin{flalign}
	\mathcal{N}(\mathbf{x}|\mathbf{m}_{1},\boldsymbol{\Sigma} _{1})&\mathcal{N}(\mathbf{x}|\mathbf{m}_{2},\boldsymbol{\Sigma}_{2})\propto \mathcal{N}(\mathbf{x}|\mathbf{m}_{c},\boldsymbol{\Sigma}_{c})&
	\\
	\boldsymbol{\Sigma}_{c} &= (\boldsymbol{\Sigma}_{1}^{-1} + \boldsymbol{\Sigma}_{2}^{-1})^{-1}  & \\
	\mathbf{m}_{c} &= \boldsymbol{\Sigma}_{c} (\boldsymbol{\Sigma}_{1}^{-1} \mathbf{m}_{1} + \boldsymbol{\Sigma}_{2}^{-1}\mathbf{m}_{2}) & 
\end{flalign}
Using the above result we can update the parameters $ \hat{\mathbf{\Theta}}_{k}^{+}$ and $\mathbf{\Sigma}_{k}^{+}$ using \eqref{eqn_vb_35}-\eqref{eqn_vb_36}.

\ifCLASSOPTIONcaptionsoff
  \newpage
\fi



%
\bibliography{main.bib}{}
\bibliographystyle{IEEEtran}




%

%
%
%




\end{document}